\documentclass[epss, 11pt]{article}
\input{psfig.sty}
\usepackage{amsmath}
\usepackage{xspace}
\usepackage{verbatim}
\usepackage{latexsym}
\usepackage{amssymb}
\usepackage{bm}
\usepackage[dvips]{graphics}
\usepackage{graphicx}
\usepackage{psfrag}
\usepackage{algorithm}
\usepackage{algorithmic}
\begin{document}
\renewcommand{\textfraction}{0}

\title{Discrete Denoising with Shifts \footnote{Authors are with the
Department of Electrical Engineering, Stanford University, Stanford,
CA 94305, USA. T.\ Weissman is also with the Department of
Electrical Engineering, Technion, Haifa 32000, Israel. E-mails: {\tt
\{tsmoon, tsachy\}@stanford.edu}. This work was partially supported
by NSF awards 0512140 and 0546535, and by a Samsung scholarship.}}
\author{
Taesup Moon  \and Tsachy Weissman }
\maketitle \thispagestyle{empty}
\begin{abstract}
We introduce S-DUDE, a new algorithm for denoising DMC-corrupted
data. The algorithm, which generalizes the recently introduced DUDE
(Discrete Universal DEnoiser) of Weissman et al., aims to compete
with a genie that has access, in addition to the noisy data, also to
the underlying clean data, and can choose to switch, up to $m$
times, between sliding window denoisers in a way that minimizes the
overall loss. When the underlying data form an individual sequence,
we show that the S-DUDE performs essentially as well as this genie,
provided that $m$ is sub-linear in the size of the data. When the
clean data is emitted by a piecewise stationary process, we show
that the S-DUDE achieves the optimum distribution-dependent
performance, provided that the same sub-linearity condition is
imposed on the number of switches. To further substantiate the
universal optimality of the S-DUDE, we show that when the number of
switches is allowed to grow linearly with the size of the data,
\emph{any} (sequence of) scheme(s) fails to compete in the above
senses. Using dynamic programming, we derive an efficient
implementation of the S-DUDE, which has  complexity (time and
memory) growing only linearly with the data size and the number of
switches $m$. Preliminary experimental results are presented,
suggesting that S-DUDE has the capacity to significantly improve on
the performance attained by the original DUDE in applications where
the nature of the data abruptly changes in time
(or  space), as is often the case in practice.\\

\end{abstract}

\normalsize

\newtheorem{claim}{Claim}
\newtheorem{guess}{Conjecture}
\newtheorem{defn}{Definition}
\newtheorem{fact}{Fact}
\newtheorem{assumption}{Assumption}
\newtheorem{theorem}{Theorem}
\newtheorem{lem}{Lemma}
\newtheorem{cor}{Corollary}
\newtheorem{proof}{Proof}
\newtheorem{pfth}{Proof of Theorem}
\newtheorem{ctheorem}{Corrected Theorem}
\newtheorem{corollary}{Corollary}
\newtheorem{proposition}{Proposition}
\newtheorem{example}{Example}
\newcommand{\mat}[2]{\ensuremath{
\left( \begin{array}{c} #1 \\ #2 \end{array} \right)}}
\newcommand{\ba}{\begin{align}
\end{align}}
\newcommand{\eq}[1]{(\ref{#1})}
\newcommand{\one}[1]{\ensuremath{\mathbf{1}_{#1}}}
\newcommand{\am}{\mbox{argmin}}
\newcommand{\dmin}{d_{\mbox{min}}}
\newcommand{\be}{\begin{eqnarray}}
\newcommand{\ee}{\end{eqnarray}}
\newcommand{\eps}{\varepsilon}
\newcommand{\imipi}{\int_{-\infty}^{\infty}}
\newcommand{\mug}{\stackrel{\triangle}{=}}
\renewcommand{\thesubsection}{\Alph{subsection}}
\def \bfpi  {\bm{\pi}}
\def \bflambda  {\bm{\lambda}}

\newcommand{\Shat}{\hat{\mathbf{S}}}
\newcommand{\Ss}{\mathbf{S}^{*}}
\newcommand{\Sb}{\mathbf{S}}
\newcommand{\Skhat}{\hat{\mathbf{S}}_{k,m}}
\newcommand{\Sks}{\mathbf{S}_k^{*}}
\newcommand{\Skb}{\mathbf{S}_k}
\newcommand{\Snm}{\mathcal{S}_{0,m}^n}
\newcommand{\Sknm}{\mathcal{S}_{k,m}^n}
\newcommand{\tL}{\tilde{L}}
\newcommand{\mcA}{\mathcal{A}}
\newcommand{\mcB}{\mathcal{B}}
\newcommand{\mcS}{\mathcal{S}}
\newcommand{\mcX}{\mathcal{X}}
\newcommand{\Xb}{\mathbf{X}}
\newcommand{\mcXhat}{\hat{\mathcal{X}}}
\newcommand{\mcZ}{\mathcal{Z}}
\newcommand{\mcN}{\mathcal{N}}
\newcommand{\cb}{\mathbf{c}}
\newcommand{\Cb}{\mathbf{C}}
\newcommand{\integers}{\mathbb{Z}}
\newcommand{\naturals}{\mathbb{N}}
\newcommand{\Vmn}{V_{m\times n}}
\newcommand{\zmn}{z_{m\times n}}
\newcommand{\mtn}{m\times n}
\newcommand{\mtnn}{m{\times} n}  
\newcommand{\mbPi}{\mathbf{\Pi}}
\newcommand{\mbm}{\mathbf{m}}
\newcommand{\mbz}{\mathbf{z}}
\newcommand{\mbu}{\mathbf{u}}
\newcommand{\xhmn}{\hat{x}_{\mtnn}}
\newcommand{\xmn}{x_{\mtnn}}
\newcommand{\Xmnuniv}{\hat{X}^{\mtnn}_{{\scriptsize{\sf univ}}}}
\newcommand{\mexp}[2]{#1{\cdot} 10^{#2}}
\renewcommand{\thesubsection}{\thesection-\Alph{subsection}}

\section{Introduction}

Discrete denoising is the problem of reconstructing the components
of a finite-alphabet sequence  based on the \emph{entire}
observation of its Discrete Memoryless Channel (DMC)-corrupted
version.  The quality of the reconstruction is evaluated via a
user-specified (single-letter) loss function. \emph{Universal}
discrete denoising, in which no statistical or other properties are
known a priori about the underlying clean data and  the goal is to
attain optimum performance, was considered and solved in
\cite{Dude}. The main problem setting there is the
``semi-stochastic'' one, in which the underlying signal is assumed
to be an ``individual sequence,'' and the randomness is due solely
to the channel noise. In this setting, it is unreasonable to expect
to attain the best performance among all the denoisers in the world,
since for every given sequence, there exists a denoiser that
recovers all the sequence components perfectly. Thus, \cite{Dude}
limits the comparison class, a.k.a. expert class, and uses the
competitive analysis approach. Specifically, it is shown that
regardless of what the underlying individual sequence may be, the
Discrete Universal DEnoiser (DUDE) essentially attains the
performance of the best sliding window denoiser that would be chosen
by a genie with access to the underlying clean sequence, in addition
to the observed noisy sequence. This semi-stochastic setting result
is shown in \cite{Dude}  to  imply the stochastic setting result,
i.e., that for any underlying stationary signal, the DUDE attains
the optimal distribution-dependent performance.

The setting of an arbitrary individual sequence, combined with
competitive analysis, has been very popular in many other research
areas, especially for problems of sequential decision-making.
Examples include universal compression \cite{LZ}, universal
prediction \cite{MF}, universal filtering \cite{UFP06}, repeated
game playing \cite{Blackwell1,Blackwell2,Hannan}, universal
portfolios \cite{CO98}, online learning
\cite{LittlestoneWarmuth,Vovk}, zero-delay coding
\cite{zerodelay,matloub}, and much more.
 A comprehensive account of this line of research can be found in \cite{CBL06}. The beauty of
this approach is the fact that it leads to the construction of
schemes that perform, on every individual sequence, essentially as
well as the best in a class of experts, which is the performance of
a genie that had hindsight on the entire sequence before selecting
his actions.  Moreover, if the expert class is judiciously chosen,
the relative sense of such a performance guarantees can, in many
cases, imply optimum performance in absolute senses as well.

One extension to this approach is competition with an expert class
and a genie that has the freedom to form a \emph{compound} action,
which breaks the sequence into a certain (limited) number of
segments, applies different experts in each segment, and achieves an
even better performance overall. Note that the optimal segmentation
of the sequence and the choice of the best expert in each segment is
also determined by hindsight. Clearly, competing with the best
compound action is more challenging, since the number of possible
compound actions is exponential in the sequence length $n$, and the
brute-force vanilla implementation of the ordinary universal scheme
requires prohibitive complexity.
 However, clever schemes with
linear complexity that successfully track the best segments and
experts have been devised in many different areas, such as online
learning, universal prediction
\cite{HerbsterWarmuth,BousquetWarmuth}, universal compression
\cite{MerhavShamir,Willems}, online linear regression \cite{A Singer
linear regression}, universal portfolios \cite{Y Singer}, and
zero-delay lossy source coding \cite{tracking the best quantizer}.


In this paper, we expand the idea of compound actions and apply it
to the discrete denoising problem. The motivation of this expansion
is natural: the characteristics of the underlying data in the
denoising problem often tend to be time- or space-varying. In this
case, determining the best segmentation and the best expert for each
segment requires complete knowledge of both clean and noisy
sequences. Therefore, whereas the challenge in sequential
decision-making problems is to track the shift of the best expert
based on the past, true observation, the challenge in the denoising
problem is to learn the shift based on the entire, but noisy,
observation. We extend DUDE to meet this challenge and provide
results that parallel and strengthen those of \cite{Dude}.

Specifically, we introduce the S-DUDE and show first that, for every
underlying noiseless sequence, it attains the performance of the
best compound finite-order sliding window denoiser (concretely
defined later), both in expectation and in a high probability sense.
We develop our scheme in the semi-stochastic setting as in
\cite{Dude}. The toolbox for the construction and analysis of our
scheme draws on ideas developed in \cite{UFP06}. We circumvent the
difficulty of not knowing the exact true loss by using an observable
 unbiased estimate of it. This kind of an estimate has proved to be
very useful in \cite{UFP06} and \cite{OWW} to devise schemes for
filtering and for denoising with dynamic contexts. Building on this
semi-stochastic setting result, we also establish a stochastic
setting result, which can be thought of as a generalization and
strengthening of the stochastic setting results of \cite{Dude}, from
the world of stationary processes to that of piecewise stationary
processes.

Our stochastic setting  has connections to other areas, such as
change-point detection problems in statistics \cite{Sieg89, Sieg95}
and switching linear dynamical systems in machine learning and
signal processing \cite{Oh06,Mesot07}. Both of these lines of
research share a common approach with S-DUDE, in that they try to
learn the change of the underlying time-varying parameter or state
of stochastic models, based on  noisy observations of the parameter
or state. One difference is that, whereas our goal is the noncausal
estimation, i.e., denoising, of the general underlying piecewise
stationary process, the change-point detection problems mainly focus
on sequentially detecting the time point where the change of model
happened. Another difference is in that the switching linear
dynamical systems focus on a special class of underlying processes,
the linear dynamical system. In addition, they  deal with
continuous-valued signals, whereas our focus is the discrete case,
with finite-alphabet signals.

As we explain in detail, the S-DUDE can be practically implemented
using a two-pass algorithm with complexity (both space and time)
linear in the sequence length and the number of switches. We also
present initial experimental results that demonstrate the S-DUDE's
potential to outperform the DUDE on both simulated and real data.

The remainder of the paper is organized as follows. Section \ref{sec
notation} provides the notation, preliminaries and background for
the paper; in Section \ref{sec sdude} we present our scheme and
establish its strong universality properties via an analysis of its
performance in the semi-stochastic setting. Section \ref{sec
stochastic} establishes the universality of our scheme in a fully
stochastic setting, where the underlying noiseless sequence is
emitted by a piecewise stationary process. Algorithmic aspects and
complexity of the actual implementation of the scheme is considered
in Section \ref{sec algorithm}, and some experimental results are
displayed in Section \ref{sec exp}. In Section \ref{sec conclusion}
we conclude with a summary of our findings and some possible future
research directions.

\section{Notation, Preliminaries, and Motivation}\label{sec notation}
\subsection{Notation}

We use a combination of notation of \cite{Dude} and \cite{UFP06}.
Let $\mcX,\mcZ,\mcXhat$ denote, respectively, the alphabet of the
clean, noisy, and reconstructed sources, which are assumed to be
finite. As in \cite{Dude} and \cite{UFP06}, the noisy sequence is a
DMC-corrupted version of the clean one, where the channel matrix
$\mathbf{\Pi}=\{\Pi(x,z)\}_{x\in\mcX,z\in\mcZ}$, $\Pi(x,z)$ denoting
the probability of a noisy symbol $z$ when the underlying clean
symbol is $x$, is assumed to be \textit{known and fixed} throughout
the paper, and \textit{of full row rank.} The $z$-th column of
$\mathbf{\Pi}$ will be denoted as $\pi_z$. Upper case letters will
denote random variables as usual; lower case letters will denote
either individual deterministic quantities or specific realizations
of random variables.

Without loss of generality, the elements of any finite set
$\mathcal{V}$ will be identified with
$\{0,1,\cdots,|\mathcal{V}|-1\}$. We let $\mathcal{V}^{\infty}$
denote the set of one-sided infinite sequences with
$\mathcal{V}$-valued components, i.e.,
$\mathbf{v}\in\mathcal{V}^\infty$ is of the form
$\mathbf{v}=(v_1,v_2,\cdots),v_i\in\mathcal{V},i\geq 1$. For
$\mathbf{v}\in\mathcal{V}^{\infty}$, let $v^n=(v_1,\cdots,v_n)$ and
$v_m^n=(v_m,\cdots,v_n)$. Furthermore, we let $v^{n\backslash t}$
denote the sequence $v^{t-1}v_{t+1}^n$. $\mathbb{R}^{\mathcal{V}}$
is a space of $|\mathcal{V}|$-dimensional column vectors with
real-valued components indexed by the elements of $\mathcal{V}$. The
$a$-th component of
$q \in \mathbb{R}^{\mathcal{V}}$ will be denoted 
by either $q_a$ or $q[a]$.
Subscripting a vector or a matrix by ``max'' will represent the
difference between the maximum and minimum of all its components.
Thus, for example, if $\Gamma$ is a $|\mcZ|\times|\mcX|$ matrix,
then $\Gamma_{\max}$ stands for
$\max_{x\in\mcX,z\in\mcZ}\Gamma(z,x)-\min_{x\in\mcX,z\in\mcZ}\Gamma(z,x)$
(in particular, if the components of $\Gamma$ are nonnegative and
$\Gamma(z,x)=0$ for some $z$ and $x$, then
$\Gamma_{\max}=\max_{z\in\mcZ, z\in\mcX}\Gamma(z,x)$.) In addition,
$\mathbf{1}_{\{\cdot\}}$ denotes an indicator of the event inside
$\{\cdot\}$.

Generally, let the finite sets $\mathcal{Y}$, $\mathcal{A}$ be,
respectively, a source alphabet and an action space. For a general
loss function $l:\mathcal{Y}\times\mathcal{A}\rightarrow\mathbb{R}$,
a \textit{Bayes response} for $\zeta\in\mathbb{R}^{\mathcal{Y}}$
under the loss function $l$ is given as
\begin{equation}\label{eq: Bayes response defined}
b_{l}(\zeta)=\arg\min_{a\in\mathcal{A}}\zeta^T\cdot\mathcal{L}_a,
\end{equation}
where $\mathcal{L}_a$ denotes the column of the matrix of the loss
function $l$ corresponding to the $a$-th action, and ties are
resolved lexicographically. The corresponding \textit{Bayes
envelope} is denoted as
\begin{eqnarray}
U_{l}(\zeta)=\min_{a\in\mathcal{A}}\zeta^T\cdot\mathcal{L}_a.\label{U
def}
\end{eqnarray}
%
Note that when $\zeta$ is a probability, namely, it has non-negative
components summing to one, $U_{l}(\zeta)$ is the minimum achievable
expected loss (as measured under the loss function $l$) in guessing
the value of $Y \in \mathcal{Y}$ which is distributed according to
$\zeta$. The associated optimal guess is $b_{l}(\zeta)$.

An \emph{$n$-block denoiser} is a collection of $n$ mappings
$\hat{\mathbf{X}}^n=\{\hat{X}_t\}_{1\leq t \leq n}$, where
$\hat{X}_t:\mcZ^n\rightarrow \mcXhat$. We assume a given loss
function $\Lambda:\mcX\times\mcXhat\rightarrow[0,\infty)$, where the
maximum single-letter loss is denoted by $\Lambda_{\max}$, and
$\lambda_{\hat{x}}$ denotes the $\hat{x}$-th column of the loss
matrix. The normalized cumulative loss of the denoiser
$\hat{\mathbf{X}}^n$ on the individual sequence pair $(x^n,z^n)$ is
represented as
$$
L_{\hat{\mathbf{X}}^n}(x^n,z^n)=\frac{1}{n}\sum_{t=1}^n\Lambda(x_t,\hat{X}_t(z^n)).
$$
In words, $L_{\hat{\mathbf{X}}^n}(x^n,z^n)$ is the normalized
(per-symbol) loss, as measured under the loss function $\Lambda$,
when using the denoiser $\hat{\mathbf{X}}^n$ and when the observed
noisy sequence is $z^n$ while the underlying clean one is $x^n$. The
notation $L_{\hat{\mathbf{X}}^n}$ is extended for $1\leq i\leq j\leq
n$,
$$
L_{\hat{\mathbf{X}}^n}(x_i^j,z^n)=\frac{1}{j-i+1}\sum_{t=i}^j\Lambda(x_t,\hat{X}_t(z^n))
$$
denoting the normalized (per-symbol) loss between (and including)
locations $i$ and $j$.

Now, consider the set $\mcS=\{s:\mcZ\rightarrow\mcXhat\}$, which is
the (finite) set of mappings that take $\mcZ$ into $\mcXhat$. We
refer to elements of  $\mcS$ as ``single-symbol denoisers'', since
each $s \in \mcS$ can be thought of as a rule for estimating $X \in
\mathcal{X}$ on the basis of $Z \in \mathcal{Z}$. Now, for any
$s\in\mcS$, an unbiased estimator for $\Lambda(x,s(Z))$ (based on
$Z$ only), where $x$ is a deterministic symbol and $Z$ is the output
of the DMC when the input is $x$, can be obtained as in
\cite{UFP06}. First, pick a function
$h:\mcZ\rightarrow\mathbb{R}^{\mcX}$ with the property that, for
$a,b\in\mcX$,
\begin{eqnarray}
E_ah_b(Z)&=&\sum_{z\in\mcZ}h_b(z)\Pi(a,z)\nonumber\\
         &=&\delta(a,b)\triangleq \left\{
\begin{array}{cc}
1, & \textrm{if $a=b$} \\
0, & \textrm{otherwise}\\
\end{array}\right\},\label{h def}
\end{eqnarray}
where $E_a$ denotes expectation over the channel output $Z$  when
the underlying channel input is $a$, and $h_b(z)$ denotes the $b$-th
component of $h(z)$. Let $H$ denote the $|\mcZ|\times|\mcX|$ matrix
whose $z$-th row is $h^T(z)$, i.e., $H(z,b)=h_b(z)$. To see that our
assumption of a channel matrix with full row rank guarantees the
existence of such an $h$, note that \eq{h def} can equivalently be
stated in matrix form as
\begin{eqnarray}
\Pi H=I,\label{identity}
\end{eqnarray}
where $I$ is the $|\mcX|\times|\mcX|$ identity matrix. Thus, e.g.,
any $H$ of the form $H=\Gamma^T(\Pi\Gamma^T)^{-1}$, for any $\Gamma$
such that $\Pi\Gamma^T$ is invertible, satisfies \eq{identity}. In
particular, $\Gamma=\Pi$ is a valid choice ($\Pi\Pi^T$ is
invertible, since $\Pi$ is of full row rank) corresponding to the
Moore-Penrose generalized inverse \cite{LT85}.
Now, for any $s\in\mcS$, $\rho(s)\in\mathbb{R}^{\mcX}$ denotes the
column vector with $x$-th component
\begin{eqnarray}
\rho_x(s)&=&\sum_{z}\Lambda(x,s(z))\Pi(x,z)\nonumber\\
         &=&E_x\Lambda(x,s(Z)) . \label{rho}
\end{eqnarray}
In words, $\rho_x(s)$ is the expected loss using the single-symbol
denoiser $s$, while the underlying symbol is $x$.
 Considering $\mcS$ as an action
space alphabet, we define a loss function
$\ell:\mcZ\times\mcS\rightarrow\mathbb{R}$ as \be
\ell(z,s)=h(z)^T\cdot\rho(s). \label{est loss} \ee We observe from
\eq{h def} and \eq{rho} that $\ell(Z,s)$ is an unbiased estimate of
$\Lambda(x,s(Z))$ since
\begin{eqnarray}
E_x\ell(Z,s)= E_x h(Z)^T\cdot\rho(s) = \sum_{x'} E_x h_{x'}(Z)
\rho_{x'}(s) = \sum_{x'} \delta (x ,  x') \rho_{x'}(s) = \rho_{x}(s)
=  E_x\Lambda(x,s(Z))\quad \ \ \forall x\in\mcX.\label{unbias}
\end{eqnarray}

For $\xi\in\mathbb{R}^{\mcZ}$, let $B_H(\xi,\cdot)\in\mcS$ be
defined by
\begin{eqnarray}
B_H(\xi,z)=\arg\min_{\hat{x}}\xi^T\cdot
H\cdot[\mathbf{\lambda}_{\hat{x}}\odot\mathbf{\pi}_z],\label{BH def}
\end{eqnarray}
where, for vectors $v_1$ and $v_2$ of equal dimensions, $v_1\odot
v_2$ denotes the vector obtained by component-wise multiplication.
Note that, similarly as in \cite[(88),(89)   ]{UFP06},
\begin{eqnarray}
B_H(\xi,\cdot)&=&\arg\min_{s\in\mcS}\sum_{z}\xi^T\cdot
H\cdot[\lambda_{s(z)}\odot\pi_z]\nonumber\\
              &=&\arg\min_{s\in\mcS}\xi^T\cdot H\cdot \rho(s)\nonumber\\
              &=&\arg\min_{s\in\mcS}\sum_z\xi_z\cdot[h^T(z)\cdot\rho(s)]\nonumber\\
              &=&\arg\min_{s\in\mcS}\sum_z\xi_z\cdot \ell(z,s)=b_{\ell}(\xi) . \label{BH map}
\end{eqnarray}
 Thus, $B_H(\xi,\cdot)$ is a Bayes response for $\xi$ under the loss function
$\ell$ defined in \eq{est loss}.


\subsection{Preliminaries}\label{prelim}

In this section, we summarize the results from \cite{Dude} and
motivate the approach underlying the construction of our new class
of denoisers. Analogously as in \cite{UFP06}, the $n$-block denoiser
$\hat{\mathbf{X}}^n=\{\hat{X}_t\}_{1\leq t \leq n}$ can be
associated with $\mathbf{F}^n=\{F_t\}_{1\leq t\leq n}$, where
$F_t:\mcZ^{n\backslash t}\rightarrow \mcS$ is defined as follows:
$F_t(z^{n\backslash t},\cdot)$ is the single-symbol denoiser in
$\mcS$ satisfying
\begin{align}
\hat{X}_t(z^n)=F_t(z^{n\backslash t},z_t) \ \ \ \ \forall
z_t.\label{F map}
\end{align}
Therefore, we can adopt the view that at each time $t$, an $n$-block
denoiser is choosing a single-symbol denoiser based on all the noisy
sequence components but $z_t$, and applying that single-symbol
denoiser on $z_t$ to yield the $t$-th reconstruction $\hat{x}_t$.
Conversely, any sequence of mappings into single-symbol denoisers
$\mathbf{F}^n$ defines a denoiser $\hat{\mathbf{X}}^n$, again via
\eq{F map}. We will adhere to this viewpoint in what follows.

One special class of widely used $n$-block denoisers is that of
$k$-th order ``sliding window'' denoisers, which we denote by
$\hat{\mathbf{X}}^{n,\mcS_k}$. Such denoisers are of the form \be
\hat{X}_t^{s_k}(z^n)=s_k(z_{t-k}^{t+k}),\qquad\quad
t=k+1,\cdots,n-k,\label{sliding def} \ee where $s_k$ is an element
of $\mcS_k=\{s_k:\mcZ^{2k+1}\rightarrow\mcXhat\}$, the (finite) set
of mappings from $\mcZ^{2k+1}$ into $\mcXhat$.\footnote{The value of
$\hat{X}_t^{s_k}(z^n)$ for $t\leq k$ and $t>n-k$ is defined, for
concreteness and simplicity, as an arbitrary fixed symbol in
$\hat{\mathcal{X}}$.} We also refer to $s_k\in\mcS_k$ as a ``$k$-th
order denoiser''. Note that $\mcS_0=\mcS$. From the definition
\eq{sliding def}, it follows that
\begin{align}
\hat{X}_i^{s_k}(z^n)=\hat{X}_j^{s_k}(z^n) \quad\textrm{whenever
$z_{i-k}^{i+k}=z_{j-k}^{j+k}$.}\label{sliding}
\end{align}
Following the association in \eq{F map}, we can adopt an alternative
view that the $k$-th order sliding window denoiser chooses a
single-symbol denoiser
$s_k(z_{t-k}^{t-1},z_{t+1}^{t+k},\cdot)\in\mcS$ at time $t$ on the
basis of the context, and
$\hat{X}_t^{s_k}(z^n)=s_k(z_{t-k}^{t-1},z_{t+1}^{t+k},z_t)$.

We denote $\cb_t\triangleq(z_{t-k}^{t-1},z_{t+1}^{t+k})$ as a
(two-sided) context for $z_t$, and define the set of all possible
$k$-th order contexts,
$\mathbf{C}_k\triangleq\{(u_{-k}^{-1},u_{1}^{k}):(u_{-k}^{-1},u_{1}^{k})\in\mcZ^{2k}\}$.
Then, for given $z^n$ and for each $\mathbf{c}\in\mathbf{C}_k$, we
define \be
\mathcal{T}(\mathbf{c})\triangleq\big\{t:\mathbf{c}_t=\mathbf{c},
\quad k+1\leq t\leq
n-k\big\}=\{t:(z_{t-k}^{t-1},z_{t+1}^{t+k})=\mathbf{c}, \quad
k+1\leq t\leq n-k\big\},\label{subseq time} \ee the set of indices
where the context equals $\mathbf{c}$. Now, an equivalent
interpretation for \eq{sliding} is that for each
$\mathbf{c}\in\mathbf{C}_k$,  the $k$-th order sliding window
denoiser employs a time-invariant single-symbol denoiser,
$s_k(\cb,\cdot)$, at all points $t\in\mathcal{T}(\mathbf{c})$. In
other words, the sequence $z^n$ is partitioned into the subsequences
associated with the various contexts, and on each such subsequence a
time-invariant single-symbol scheme is employed.

In \cite{Dude}, for integers $k\geq 0$ and $n>2k$, the $k$-th order
minimum loss of $(x^n,z^n)$ is defined by
\begin{eqnarray}
D_k(x^n,z^n)&\triangleq&
\min_{\hat{\mathbf{X}}^n\in\hat{\mathbf{X}}^{n,\mcS_k}}L_{\hat{\mathbf{X}}^n}(x_{k+1}^{n-k},z^n)\nonumber\\
            &=&\min_{s_k\in\mcS_k}\frac{1}{n-2k}\sum_{t=k+1}^{n-k}\Lambda(x_t,s_k(\cb_t,z_t)).\label{Dmin}
\end{eqnarray}
The identity of the element $s_k\in\mcS_k$ that achieves \eq{Dmin}
depends not only on $z^n$, but also on $x^n$, since \eq{Dmin} can be
expressed as
$$\frac{1}{n-2k}\sum_{\cb\in\Cb_k}\bigg[\min_{s\in\mcS}\sum_{\tau\in\mathcal{T}(\cb)}\Lambda(x_\tau,s(z_\tau))\bigg],
$$ and at each time $t$, the best $k$-th order sliding window
denoiser that achieves \eq{Dmin} will employ the single-symbol
denoiser \be
\arg\min_{s\in\mcS}\sum_{\tau\in\mathcal{T}(\cb_t)}\Lambda(x_\tau,s(z_\tau)),\label{argmin
true} \ee which is determined from the joint empirical distribution
of pairs $\{(x_\tau,z_\tau):\tau\in\mathcal{T}(\cb_t)\}$.

It was shown in \cite{Dude} that, despite the lack of knowledge of
$x^n$, $D_k(x^n,Z^n)$ is achievable in a sense made precise below,
in the limit of growing $n$, by a scheme that only has access to
$Z^n$. This scheme is dubbed in \cite{Dude} as the Discrete
Universal DEnoiser (DUDE), $\hat{\mathbf{X}}^{n,k}_{\mathrm{univ}}$.
The algorithm is defined by
\begin{eqnarray}
\hat{X}^k_{\mathrm{univ},t}(z^n)&=&B_H(\mathbf{m}(z^n,z_{t-k}^{t-1},z_{t+1}^{t+k}),z_t),\label{dude}
\end{eqnarray}
where $\mathbf{m}(z^n, \mathbf{c})$ is the vector of counts of the
appearances of the various symbols within the context $\mathbf{c}$
along the sequence $z^n$. That is, for all $\beta\in\mcZ$,
$\mathbf{m}(z^n,\tilde{z}_{-k}^{-1},\tilde{z}_{1}^{k})$ is the
$|\mathcal{Z}|$-dimensional column vector whose $\beta$-th component
is
$$
\mathbf{m}(z^n,\tilde{z}_{-k}^{-1},\tilde{z}_{1}^{k})[\beta]=\big|\{t:k+1\leq
t\leq n-k, z_{t-k}^{t+k}=\tilde{z}_{-k}^{-1}\beta
\tilde{z}_{1}^{k}\}\big|,
$$
namely,  the number of appearances of $\tilde{z}_{-k}^{-1}\beta
\tilde{z}_{1}^{k}$ along the sequence $z^n$.

The main result of \cite{Dude} is the following theorem, pertaining
to the semi-stochastic setting of  an individual sequence
$\mathbf{x} = (x_1, x_2, \ldots )$ corrupted by a DMC that yields
the stochastic noisy sequence $\mathbf{Z} = (Z_1, Z_2, \ldots)$.
\begin{theorem}(\cite[Theorem 1]{Dude})\label{dude main thm}
Take $k=k_n$ satisfying $k_n|\mcZ|^{2k_n}=o(n/\log n)$. Then, for
all $\mathbf{x}\in\mcX^{\infty}$, the sequence of denoisers
$\{\hat{\mathbf{X}}^{n,k_n}_{\mathrm{univ}}\}$ defined in \eq{dude}
satisfies:
\begin{enumerate}
\item[a)]
$$\lim_{n\rightarrow\infty}\Big[L_{\hat{\mathbf{X}}^{n,k_n}_{\mathrm{univ}}}(x^n,Z^n)-D_{k_n}(x^n,Z^n)\Big]=0\quad\textrm{a.s.}$$
\item[b)]
$$
E\Big[L_{\hat{\mathbf{X}}^{n,k_n}_{\mathrm{univ}}}(x^n,Z^n)-D_{k_n}(x^n,Z^n)\Big]=O\left(\sqrt{\frac{k_n|\mcZ|^{2k_n}}{n}}\right).
$$
\end{enumerate}
\end{theorem}
Theorem \ref{dude main thm} was further shown in \cite{Dude} to
imply the universality of the DUDE in the fully stochastic setting
where the underlying sequence is emitted by a stationary source (and
the goal is to attain the performance of the optimal
distribution-dependent denoiser).

From \eq{dude}, it is apparent that the DUDE ends up employing a
$k$-th order sliding window denoiser (where the sliding window
scheme the DUDE chooses depends on $z^n$). Moreover, \eq{BH map}
implies that, at each time $t$, DUDE is merely employing the
single-symbol denoiser
$B_H(\mathbf{m}(z^n,z_{t-k}^{t-1},z_{t+1}^{t+k}),\cdot)\in\mcS$,
which can be obtained by finding the Bayes response
$b_{\ell}\big(\mathbf{m}(z^n,z_{t-k}^{t-1},z_{t+1}^{t+k})\big)$ or,
equivalently, the mapping in $\mcS$ given by \be
\arg\min_{s\in\mcS}\sum_{\tau\in\mathcal{T}(\cb_t)}\ell(z_\tau,s),\label{argmin
est}\ee where $\ell(z,s)$ is the loss function defined in \eq{est
loss}. By comparing \eq{argmin true} with \eq{argmin est}, and from
Theorem \ref{dude main thm}, we observe that working with the
estimated loss $\ell(z_{\tau},s)$ in lieu of the genie-aided
$\Lambda(x_{\tau},s(z_{\tau}))$ allows us to essentially achieve the
genie-aided performance in \eq{Dmin}.

\subsection{Motivation}
Our motivation for this paper is based on the observation that the
$k$-th order sliding window denoisers ignore the time-varying nature
of the underlying sequence $x^n$. That is, as discussed above, for
time instances with the same contexts, the single-symbol denoiser
employed along the associated subsequence is time-invariant. In
other words, for each $t$, only the empirical distribution of the
sequence $\{(x_\tau,z_\tau):\tau\in\mathcal{T}(\cb_t)\}$ matters,
but its order of composition, i.e., its time-varying nature, is not
considered. It is clear, however, that when the characteristics of
the underlying clean sequence $x^n$ are changing, the (normalized)
cumulative loss that is achieved by sliding window denoisers that
can shift from one rule to another along the sequence may be
strictly lower (better) than \eq{Dmin}. We now devise and analyze
our new scheme that achieves this more ambitious target performance.

\section{The Shifting Denoiser (S-DUDE)}\label{sec sdude}

In this section, we derive our new class of denoisers and analyze
their performance. In Subsection \ref{sbs section}, we begin with
the simplest case, competing with shifting symbol-by-symbol
denoisers, or, in other words, shifting $0$-th order denoisers. The
argument is generalized to shifting $k$-th order denoisers in
Subsection \ref{kth order section}, and the framework and results
include Subsection \ref{sbs section} as a special case. We will use
the notation $\mcS_0$, instead of $\mcS$, for consistency in
denoting the class of single-symbol denoisers. Throughout this
section, we assume the semi-stochastic setting.

\subsection{Switching between symbol-by-symbol ($0$-th order) denoisers}\label{sbs section}
Consider an $n$-tuple of single-symbol denoisers
$\mathbf{S}=\{s_1,\cdots,s_n\}\in\mcS_0^n$. Then, as mentioned in
Section \ref{prelim}, for such $\mathbf{S}$, we can define the
associated $n$-block denoiser $\hat{\mathbf{X}}^{n,\mathbf{S}}$ as
\begin{eqnarray}
\hat{X}^{\mathbf{S}}_t(z^n)=s_t(z_t).\label{XS m shift}
\end{eqnarray}
Note that in this case, the single-symbol denoiser applied at each
time may depend on the time $t$ (but not on $z^{n\backslash t}$, as
would be the case for a general denoiser). We also denote the
estimated normalized cumulative loss as
 \begin{eqnarray}
\tL_\mathbf{S}(z^n) &\triangleq&
\frac{1}{n}\sum_{t=1}^n\ell(z_t,s_t),\label{est avg loss}
\end{eqnarray}
whose property is given in the following lemma, which parallels
\cite[Theorem 4]{UFP06}.

\begin{lem}\label{lem from martingale}
Fix $\epsilon>0$. For fixed $\mathbf{S}\in\mcS_0^n$, and all
$x^n\in\mcX^{n}$,
\begin{eqnarray}
P\Big(L_{\hat{\mathbf{X}}^{n,\mathbf{S}}}(x^n,Z^n)-\tL_{\mathbf{S}}(Z^n)>\epsilon\Big)&\leq&\exp\Big(-n\frac{2\epsilon^2}{L_{\max}^2}\Big)\quad\textrm{and}\label{eq:lemma 1 eq1}\\
P\Big(\tL_{\mathbf{S}}(Z^n)-L_{\hat{\mathbf{X}}^{n,\mathbf{S}}}(x^n,Z^n)>\epsilon\Big)&\leq&\exp\Big(-n\frac{2\epsilon^2}{L_{\max}^2}\Big),\label{eq:lemma1
eq2}
\end{eqnarray}
where $L_{\max}=\Lambda_{\max}+\ell_{\max}$.
\end{lem}
 In words, the lemma shows that for every $\mathbf{S}\in\mcS_0^n$,
the estimated loss $\tL_{\mathbf{S}}(Z^n)$ is concentrated around
the true loss $L_{\hat{\mathbf{X}}^{n,\mathbf{S}}}(x^n,Z^n)$ with
high probability, as $n$ becomes large, regardless of the underlying
sequence $x^n$.

\emph{Proof of Lemma \ref{lem from martingale}:} See
Appendix \ref{proof of lem from martingale}.$\quad\blacksquare$\\

Now, let the integer $0\leq m\leq\lfloor \frac{n}{2}\rfloor$ denote
the maximum number of shifts allowed along the sequence. Then,
define a set $\mcS_{0,m}^n\subseteq\mcS_0^n$ as
\begin{eqnarray}
\Snm=\Big\{\mathbf{S}\in\mcS_0^n:\sum_{t=2}^n\mathbf{1}_{\{s_{t-1}\neq
s_t\}}\leq m\Big\},\label{Snm def}
\end{eqnarray}
namely, $\mcS_{0,m}^n$ is the set of $n$-tuples of single-symbol
denoisers with at most $m$ shifts from one mapping to
another.\footnote{Note that, when $m=0$, $\mcS_{0,0}^n$ is the set
of constant $n$-tuples consisting of the same single-symbol
denoiser.} Analogously to \eq{Dmin}, for the class of $n$-block
denoisers $\hat{\mathbf{X}}^{n,\mathbf{S}}$ with
$\mathbf{S}\in\Snm$, we define
\begin{eqnarray}
D_{0,m}(x^n,z^n)&\triangleq&\min_{\mathbf{S}\in\mcS_{0,m}^n}L_{\hat{\mathbf{X}}^{n,\mathbf{S}}}(x^n,z^n)\nonumber\\
                &=&\min_{\mathbf{S}\in\Snm}\frac{1}{n}\sum_{t=1}^n\Lambda(x_t,s_t(z_t)),\label{D0min}
\end{eqnarray}
which is the minimum normalized cumulative loss that can be achieved
for $(x^n,z^n)$ by the sequence of $n$ single-symbol denoisers that
allow at most $m$ shifts. Our goal in this section is to build a
universal scheme that only has access to $Z^n$, but still
essentially achieves $D_{0,m}(x^n,Z^n)$.

As hinted by the DUDE, we build our universal scheme by working with
the estimated loss. That is, define
\begin{eqnarray}
\hat{\mathbf{S}} = \hat{\mathbf{S}}
(z^n)&\triangleq&\arg\min_{\Sb\in\mathcal{S}_{0,m}^n}\tilde{L}_\mathbf{S}(z^n),\label{Shat
def}
\end{eqnarray}
and our $(0,m)$-Shifting Discrete Universal DEnoiser (S-DUDE),
$\hat{\mathbf{X}}^{n,0,m}_{\mathrm{univ}}$, is defined as
$\hat{\mathbf{X}}^{n,\hat{\mathbf{S}}}$. It is clear that, by
definition, $L_{\hat{\Xb}^{n,\Shat}}(x^n,z^n)\geq D_{0,m}(x^n,z^n)$
for all $x^n$ and $z^n$, but we can also show that, with high
probability, $L_{\hat{\Xb}^{n,\Shat}}(x^n,Z^n)$ does not exceed
$D_{0,m}(x^n,Z^n)$ by much, as stated in the following theorem.

\begin{theorem}\label{concentration thm}
Let $\hat{\mathbf{X}}^{n,0,m}_{\mathrm{univ}}$ be defined as
$\hat{\mathbf{X}}^{n,\hat{\mathbf{S}}}$, where $\Shat$ is given in
\eq{Shat def}. Then, for all  $\epsilon>0$ and $x^n\in\mcX^n$,
$$
P\Big(L_{\hat{\mathbf{X}}^{n,0,m}_{\mathrm{univ}}}(x^n,Z^n)-D_{0,m}(x^n,Z^n)>\epsilon\Big)\leq
2\exp\left(-n\Big[\frac{\epsilon^2}{2L_{\max}^2}-2\Big\{h\Big(\frac{m}{n}\Big)+\frac{(m+1)\ln
N}{n}\Big\}\Big]\right),
$$
where $h(x)=-x\ln x-(1-x)\ln (1-x)$ for $0\leq x\leq1$, and
$N=|\mcS|=|\mcZ|^{|\mcXhat|}$.
In particular, the right-hand side of the inequality is
exponentially small, provided $m=o(n)$.
\end{theorem}
\emph{Remark:} It is reasonable to expect this theorem to hold,
given Lemma \ref{lem from martingale}. That is, since, for fixed
$\mathbf{S}\in\mcS_{0,m}^n$, $\tilde{L}_{\Sb}(Z^n)$ is concentrated
on $L_{\hat{\mathbf{X}}^{n,\Sb}}(x^n,Z^n)$, it is plausible that
$\hat{\Sb}$ that achieves
$\min_{\Sb\in\mcS_{0,m}^n}\tilde{L}_{\Sb}(Z^n)$ will have a loss
$L_{\hat{\mathbf{X}}^{n,\hat{\Sb}}}(x^n,Z^n)$ close to
$\min_{\Sb\in\mcS_{0,m}^n}L_{\hat{\mathbf{X}}^{n,\Sb}}(x^n,Z^n)$,
i.e., $D_{0,m}(x^n,Z^n)$.

\emph{Proof of Theorem \ref{concentration thm}:} See Appendix
\ref{proof of concentration thm}.$\quad\blacksquare$
\subsection{Switching between $k$-th order denoisers}\label{kth
order section} Now, we extend the result from Subsection \ref{sbs
section} to the case of shifting between $k$-th order denoisers. The
argument parallels that of  Subsection \ref{sbs section}. Let
$\{s_{k,t}\}_{t=k+1}^{n-k}$ be an arbitrary sequence of the $k$-th
order denoiser mappings, i.e., $s_{k,t}\in\mcS_k$ for $k+1\leq t\leq
n-k$. Now, for given $z^n$, define an $(n-2k)$-tuple of ($k$-th
order denoiser induced) single-symbol denoisers
\begin{eqnarray}
\Sb_k(z^n)\triangleq\{s_{k,t}(\mathbf{c}_t,\cdot)\}_{t=k+1}^{n-k}\in\mcS_0^{n-2k},\label{Sb
def}
\end{eqnarray}
where, to recall, $\cb_t=(z_{t-k}^{t-1},z_{t+1}^{t+k})$, and
$s_{k,t}(\mathbf{c}_t,\cdot)$ is the single-symbol denoiser induced
from $s_{k,t}\in\mcS_k$ and $\mathbf{c}_t$. For brevity of notation,
we will suppress the dependence on $z^n$ in $\Sb_k(z^n)$ and denote
it as $\Sb_k$. Then, as in \eq{XS m shift}, we define the associated
$n$-block denoiser $\hat{\Xb}^{n,\Sb_k}$ as \footnote{Again, the
value of $\hat{\Xb}_t^{\Sb_k}(z^n)$ for $t\leq k$ and $t>n-k$ can be
defined as an arbitrary fixed symbol, since it will be
inconsequential in subsequent development.}
\begin{eqnarray}
\hat{\Xb}_t^{\Sb_k}(z^n)=s_{k,t}(\cb_t,z_t).\label{XSk m shift}
\end{eqnarray}
In addition, extending \eq{est avg loss}, the estimated normalized
cumulative loss is given as
\begin{eqnarray}
\tL_{\Sb_k}(z^n)=\frac{1}{n-2k}\sum_{t=k+1}^{n-k}\ell(z_t,s_{k,t}(\cb_t,\cdot)).\label{est
k avg loss}
\end{eqnarray}
Then, we have the following lemma, which parallels Lemma \ref{lem
from martingale}.
\begin{lem}\label{kl concent}
Fix $\epsilon>0$. For any fixed sequence
$\{s_{k,t}\}_{t=k+1}^{n-k}$, and all $x^n\in\mcX^n$,
\begin{eqnarray}
\mathrm{Pr}\Big(L_{\hat{\Xb}^{n,\Sb_k}}(x^{n-k}_{k+1},Z^n)-\tL_{\Sb_k}(Z^n)>\epsilon\Big)&\leq&
(k+1)\exp\Bigg(-\frac{2(n-2k)\epsilon^2}{(k+1)L_{\max}^2}\Bigg)\quad\textrm{and}\label{eq: lemma3 eq1}\\
\mathrm{Pr}\Big(\tL_{\Sb_k}(Z^n)-L_{\hat{\Xb}^{n,\Sb_k}}(x^{n-k}_{k+1},Z^n)>\epsilon\Big)&\leq&
(k+1)\exp\Bigg(-\frac{2(n-2k)\epsilon^2}{(k+1)L_{\max}^2}\Bigg),\label{eq: lemma3 eq2}
\end{eqnarray}
where $L_{\max}=\Lambda_{\max}+\ell_{\max}$.
\end{lem}
\emph{Remark:} Note that when $k=0$, this lemma coincides with Lemma
\ref{lem from martingale}. The proof of this lemma combines Lemma
\ref{lem from martingale} and the de-interleaving argument in the
proof of \cite[Theorem 2]{Dude}. Namely, we de-interleave $Z^n$ into
$(k+1)$ subsequences  consisting of symbols separated by blocks of
$k$ symbols, and exploit the conditional independence of symbols in
each subsequence, given all symbols not in that subsequence, to use
Lemma \ref{lem from martingale}.

\emph{Proof of Lemma \ref{kl concent}:} See Appendix \ref{proof of
kl concent}. $\quad\blacksquare$\\

Now, for an integer $0\leq m\leq \lfloor\frac{n-2k}{2}\rfloor$ and
given $z^n$, let $n(\cb)\triangleq|\mathcal{T}(\cb)|$, and
$m(\cb)\triangleq\min\{n(\cb),m\}$ for $\cb\in\Cb_k$. Then,
analogously as in \eq{Snm def}, we define
\begin{eqnarray}
\mcS_{k,m}^{n}(z^n)=\Big\{\Sb_k(z^n)\in\mcS_0^{n-2k}:\{s_{k,\tau}(\mathbf{c},\cdot)\}_{\tau\in\mathcal{T}(\mathbf{c})}\in\mcS_{0,m(\cb)}^{n(\mathbf{c})}\quad\textrm{for
all} \quad\mathbf{c}\in\mathbf{C}_k\Big\} . \label{Skmn def}
\end{eqnarray}
In words, $\mcS_{k,m}^{n}(z^n)$ is the set of $(n-2k)$-tuples of
($k$-th order denoiser induced) single-symbol denoisers that allow
at most $m(\cb)$ shifts within the subsequence
$\{t:t\in\mathcal{T}(\cb)\}$ for each context
$\cb\in\Cb_k$.\footnote{When $m=0$, $\mcS_{k,0}^{n}(z^n)$ becomes
the set of $n$-block $k$-th order `sliding window' denoisers.}
Again, for brevity, the dependence on $z^n$ in $\mcS_{k,m}^{n}(z^n)$
is suppressed, and we write simply $\mcS_{k,m}^{n}$. It is worth
noting that $\mcS_{k,m}^n$ is a larger class than the class of
$k$-th order `sliding window' denoisers that are allowed to shift
 at most $m$ times. The reason is that in $\mcS_{k,m}^n$,
the shift within each subsequence associated with each context can
occur at any time, regardless of the shifts in other subsequences,
whereas in the latter class, the shifts in each subsequence occur
together with other shifts in other subsequences.

For integers $k\geq 0$ and $n>2k$, we now define, for the class of
$n$-block denoisers $\hat{\Xb}^{n,\Sb}$ with $\Sb\in\mcS_{k,m}^n$,
\begin{eqnarray}
D_{k,m}(x^n,z^n)&\triangleq&\min_{\mathbf{S}\in\mcS_{k,m}^n}L_{\hat{\mathbf{X}}^{n,\mathbf{S}}}(x^{n-k}_{k+1},z^n)\nonumber\\
                &=&\min_{\mathbf{S}\in\mcS_{k,m}^n}\frac{1}{n-2k}\sum_{t=k+1}^{n-k}\Lambda(x_t,s_{k,t}(\cb_t,z_t)),\label{Dkmin}
\end{eqnarray}
the minimum normalized cumulative loss of $(x^n,z^n)$ that can be
achieved by the sequence of $k$-th order denoisers that allow at
most $m$ shifts within each context. Now, to build a legitimate (non
genie-aided) universal scheme achieving \eq{Dkmin} on the basis of
$Z^n$ only, we define
\begin{eqnarray}
\hat{\Sb}_{k,m}=\arg\min_{\Sb\in\mcS_{k,m}^n}\tL_{\Sb}(z^n),\label{Skhat
def}
\end{eqnarray}
and the $(k,m)$-S-DUDE, $\hat{\Xb}^{n,k,m}_{\mathrm{univ}}$, is
defined as $\hat{\Xb}^{n,\hat{\Sb}_{k,m}}$. Note that when $m=0$,
$\hat{\Xb}^{n,\hat{\Sb}_{k,m}}$  coincides with the DUDE in
\cite{Dude}.  The following theorem generalizes Theorem
\ref{concentration thm} to the case of general $k \geq 0$.
\begin{theorem}\label{k concentration thm}
Let $\hat{\Xb}^{n,k,m}_{\mathrm{univ}}$ be given by
$\hat{\Xb}^{n,\hat{\Sb}_{k,m}}$, where $\hat{\Sb}_{k,m}$ is defined
in \eq{Skhat def}. Then, for all $\epsilon>0$ and $x^n\in\mcX^n$,
\begin{eqnarray}
&
&\mathrm{Pr}\Big(L_{\hat{\Xb}^{n,k,m}_{\mathrm{univ}}}(x^{n-k}_{k+1},Z^n)-D_{k,m}(x^n,Z^n)>\epsilon\Big)\\
&\leq&2(k+1)\exp\Bigg(-(n-2k)\cdot\Big[\frac{\epsilon^2}{2(k+1)L_{\max}^2}-2|\mcZ|^{2k}\cdot\Big\{h\Big(\frac{m}{n-2k}\Big)+\frac{(m+1)\ln
N}{n-2k}\Big\}\Big]\Bigg) , \label{k concent part b}
\end{eqnarray}
where $h(x)=-x\ln x-(1-x)\ln (1-x)$ for $0\leq x\leq1$, and
$N=|\mcS|=|\mcZ|^{|\mcXhat|}$.
\end{theorem}
\emph{Remark:} Note that when $k=0$, this theorem coincides with
Theorem \ref{concentration thm}. Similarly to the way Theorem
\ref{concentration thm} was plausible given Lemma \ref{lem from
martingale}, Theorem \ref{k concentration thm} can be expected given
Lemma \ref{kl concent}, since $\hat{\Sb}_{k,m}$ achieves
$\min_{\Sb\in\mcS_{k,m}^n}\tL_{\Sb}(Z^n)$, and we expect
$L_{\hat{\mathbf{X}}^{n,\hat{\Sb}_{k,m}}}(x^{n-k}_{k+1},Z^n)$ to be
close to $D_{k,m}(x^n,Z^n)$ from the concentration of
$\tL_{\Sb}(Z^n)$ to
$L_{\hat{\mathbf{X}}^{n,\mathbf{S}}}(x^{n-k}_{k+1},Z^n)$ for all
$\Sb\in\mcS_{k,m}^n$.

\emph{Proof of Theorem \ref{k concentration thm}:} See Appendix
\ref{proof of k concentration thm}. $\quad\blacksquare$\\

From Theorem \ref{k concentration thm}, we now easily obtain one of
the main results of the paper, which extends Theorem 1 from the case
$m=0$ to the case of general $0\leq m \leq\lfloor
\frac{n-2k}{2}\rfloor$. That is, the following theorem asserts that,
for every underlying sequence $\mathbf{x}\in\mcX^{\infty}$, our
$(k,m)$-S-DUDE performs essentially as well as the best shifting
$k$-th order denoiser that allows at most $m$ shifts within each
context, both in high probability and expectation sense, provided a
growth condition on $k$ and $m$ is satisfied.

\begin{theorem}\label{semi stochastic theorem}
Suppose $k=k_n$ and $m=m_n$ are such that the right-hand side of \eq{k
concent part b} is summable in $n$. Then, for all $\mathbf{x}\in\mcX^{\infty}$, the sequence of
denoisers $\{\hat{\Xb}^{n,k,m}_{\mathrm{univ}}\}$ satisfies
\begin{enumerate}
\item[a)]
\begin{equation}\label{eq: main a.s. result}
    \lim_{n\rightarrow\infty}\Big[L_{\hat{\Xb}^{n,k,m}_{\mathrm{univ}}}(x^n,Z^n)-D_{k,m}(x^n,Z^n)\Big]=0\quad\mathrm{a.s.}
\end{equation}

\item[b)] For any $\delta > 0$,
\begin{equation}\label{eq: bound in main theorem on exp performance}
    E\Big[L_{\hat{\Xb}^{n,k,m}_{\mathrm{univ}}}(x^n,Z^n)-D_{k,m}(x^n,Z^n)\Big]=O\left(\sqrt{k_n|\mcZ|^{2k_n}\Big(\frac{m_n}{n}\Big)^{1-\delta}}\right)
    .
\end{equation}
%
%
%
%
\end{enumerate}
\end{theorem}
\emph{Remark:} It will be seen in Claim \ref{claim: k m condition}
below that the stipulation in the theorem implies
$\lim_{n\rightarrow\infty}k_n|\mcZ|^{2k_n}\big(\frac{m_n}{n}\big)^{1-\delta}=0$,
which, when combined with (\ref{eq: bound in main theorem on exp
performance}), implies that the expected difference on the left hand
side of (\ref{eq: bound in main theorem on exp performance})
vanishes with increasing $n$. That in itself, however, can easily be
deduced from \eq{eq: main a.s. result} and bounded convergence. The
more significant value of \eq{eq: bound in main theorem on exp
performance} is in providing a rate of convergence result for the
`redundancy' in the S-DUDE's performance, as a function of both $k$
and $m$. In particular, note that for any $\eta > 0$, $O ( n^{-1/2 +
\eta})$ is achievable provided $k_n = c \log n$ and $m_n = n^\xi$,
for small enough positive constants $c, \xi$.\\

In what follows, we specify the maximal growth rates for $k=k_n$ and
$m=m_n$ under which the summability condition stipulated  in Theorem
\ref{semi stochastic theorem} holds.
\begin{claim}\label{claim: k m condition}
\begin{enumerate}
\item[a)] Maximal growth rate for $k$: The summability condition in Theorem \ref{semi
stochastic theorem} is satisfied provided $k_n=c_1 \log n$ with
$c_1<\frac{1}{2\log |\mathcal{Z}|}$ and $m_n$ grows at any
sub-polynomial rate. On the other hand, the condition is not
satisfied for $k_n=c_1 \log n$ with any $c_1 \geq \frac{1}{2\log
|\mathcal{Z}|}$, even when $m$ is fixed (not growing with $n$).

\item[b)] Maximal  growth rate for $m$: The summability condition in Theorem \ref{semi
stochastic theorem} is satisfied for any sub-linear growth rate of
$m_n$, provided $k_n$ is taken to increase sufficiently slowly that
$k_n |\mathcal{Z}|^{2 k_n} = o((n/m_n)^{1-\delta})$ for some $\delta
> 0$.  On the other hand, the condition is not satisfied whenever
$m_n$ grows linearly with $n$,  even when $k$ is fixed.
\end{enumerate}
\end{claim}
\emph{Proof of Claim \ref{claim: k m condition}:} See Appendix
\ref{proof of claim: k m condition}. $\quad\blacksquare$\\
%
%

\emph{Proof of Theorem \ref{semi stochastic theorem}:} See Appendix
\ref{proof of semi stochastic theorem}. $\quad\blacksquare$

%

\subsection{A ``strong converse''}
In Claim \ref{claim: k m condition}, we have shown the necessity of
$m=o(n)$ for the condition required in Theorem \ref{semi stochastic
theorem} to hold. However, we can prove the necessity of $m=o(n)$ in
a much stronger sense, described in the following theorem.
\begin{theorem} \label{th: converse establishing necessity}
Suppose that $\mathcal{X} = \hat{\mathcal{X}}$, that $\Lambda (x,
\hat{x}) \geq 0$ for all $x, \hat{x}$  with equality if and only if
$x = \hat{x}$, and that $\Pi (x, z) > 0$ for all $x, z$. If $m =
\Theta (n)$, then for \underline{any} sequence of denoisers $\{
\hat{\mathbf{X}}^n \}$, there exists $\mathbf{x}^{\infty} \in
\mathcal{X}^\infty$ such that
\begin{equation}\label{eq: limsup of exp diff is positive}
    \limsup_{n \rightarrow \infty} E \left[ L_{\hat{\mathbf{X}}^n}  (x^n, Z^n) - D_{0, m} (x^n, Z^n)
    \right]  > 0.
\end{equation}
\end{theorem}
\emph{Remark: }The theorem establishes the fact that when $m = o(n)$
does not hold, namely, when $m = \Theta (n)$, not only does the
almost sure convergence in Theorem \ref{semi stochastic theorem} not
hold but, in fact, even the much weaker convergence in expectation
would fail. Further, it shows that this would be the case for
\emph{any} sequence of denoisers, not necessarily the S-DUDE.
 Furthermore, \eq{eq:
limsup of exp diff is positive} features $D_{0, m} (x^n, Z^n)$,
pertaining to competition with a genie that shifts among
single-symbol denoisers so, \emph{a fortiori}, it implies that for
any fixed $k>0$ or $k$ that grows with $n$,
\begin{equation}\label{eq: limsup of exp diff is positive for any k}
    \limsup_{n \rightarrow \infty} E \left[ L_{\hat{\mathbf{X}}^n}  (x^n, Z^n) - D_{k, m} (x^n, Z^n)
    \right]  > 0
\end{equation}
also holds since, by definition, $D_{0, m} (x^n, z^n)\geq D_{k, m}
(x^n, z^n)$ for all $x^n,z^n$ and $k \geq 0$. Therefore, the theorem
asserts that for \emph{any} sequence of denoisers to compete with
$D_{k, m} (x^n, Z^n)$, even in expectation sense, $m=o(n)$ is
necessary. Finally, we mention that the conditions stipulated in the
statement of the theorem regarding the loss function and the channel
can be considerably relaxed without compromising the validity of the
theorem. These conditions are made to allow for the simple proof
that we give in  Appendix \ref{proof of th: converse establishing
necessity}.

%

\section{The Stochastic Setting}\label{sec stochastic}

In \cite{Dude}, the semi-stochastic setting result, \cite[Theorem
1]{Dude}, was shown to imply the result for the stochastic setting
as well. That is, when the underlying data form  a stationary
process, \cite[Section VI]{Dude} shows that the DUDE attains optimum
distribution-dependent performance. Analogously, we can now use the
results from the semi-stochastic setting of the previous section to
generalize the results of \cite[Section VI]{Dude} and show that
 our S-DUDE
attains optimum distribution-dependent performance when the
underlying data form a piecewise stationary process. We first define
the precise notion of the class of piecewise stationary processes in
Subsection \ref{stochastic subsection 1}, and discuss the richness
of this class  in Subsection \ref{stochastic subsection 2}.
Subsection \ref{stochastic subsection 3} gives the main result of
this section: the stochastic setting optimality of the S-DUDE.

%

\subsection{Definition of the class of processes $\mathcal{P}
\{m_n\}$}\label{stochastic subsection 1}

Let $P_{\mathbf{X}}^{(1)},\cdots,P_{\mathbf{X}}^{(M)}$ be a finite
collection of $M$ probability distributions of stationary processes,
with components taking the values in $\mcX$. Let $\mathbf{A}$ be a
process  with components taking the values in $\{1,\ldots,M\}$.
Then, a piecewise stationary process $\mathbf{X}$ is generated by
shifting between the $M$ processes in a way specified by the
``switching process'' $\mathbf{A}$, as we now describe.

%

First, denote $r(A^n)$ as the number of shifts that have occurred
along the $n$-tuple $A^n$, i.e.,
$$
r(A^n)\triangleq \sum_{j=1}^{n-1}\mathbf{1}_{\{A_j\neq A_{j+1}\}}.
$$
Thus, there are $r(A^n)+1$ ``blocks'' in $A^n$, where each block is
a tuple of constant values that are different from the values of
adjacent blocks. Now, for each $1\leq i\leq r(A^n)+1$, we define
$$
\tau_{i}(A^n)\triangleq\left\{
\begin{array}{cc}
\inf\{t:\sum_{j=1}^t\mathbf{1}_{\{A_j\neq A_{j+1}\}}=i\} & \textrm{if\quad $1\leq i\leq r(A^n)$ } \\
n & \textrm{if \quad $i=r(A^n)+1$}\\
\end{array}\right.
$$
as the last time instance of the $i$-th block in $A^n$. In addition,
define $\tau_0(A^n) \triangleq 0$. Clearly, $r(A^n)$ and
$\tau_i(A^n)$ depend on $A^n$ and, thus, are random variables.
However, for brevity, we suppress the dependence on $A^n$ when there
is no confusion, and write simply $r$ and $\tau_i$, respectively.

Using these definitions, and by denoting $P_{A^n}$ as the $n$-th
order marginal distribution of $\mathbf{A}$, we define a piecewise
stationary process $\mathbf{X}$ by characterizing its $n$-th order
marginal distribution $P_{X^n}$ as
\begin{eqnarray}
P_{X^n}(X^n=x^n)&=&\sum_{a^n}P_{A^n}(a^n) P(X^n=x^n|A^n=a^n)\nonumber\\
           &=&\sum_{a^n}P_{A^n}(a^n)\prod_{i=1}^{r+1}P_{\mathbf{X}}^{(a_{\tau_i})}(x_{\tau_{i-1}+1}^{\tau_i}),\label{PX
def}
\end{eqnarray}
for each $n$. The corresponding distribution of the process
$\mathbf{X}$ is denoted as $P_{\mathbf{X}}$.\footnote{$\{ P_{X^n}
\}_{n \geq 1}$ is readily verified to be a consistent family of
distributions and, thus, by Kolmogorov's extension theorem, uniquely
defines the distribution of the process  $\mathbf{X}$.} In words,
$\mathbf{X}$ is constructed by following one of the $M$ probability
distributions in each block, switching from one to another depending
on $\mathbf{A}$. Furthermore, conditioned on the realization of
$\mathbf{A}$, each stationary block is independent of other blocks,
even if the distribution of distinct blocks is the same. This
property of conditional independence is reasonable for modeling many
types of data arising in practice, since we can think of the $M$
distributions as different `modes'; if the process returns to the
same mode, it is reasonable to model the new block as a new
independent realization of that same distribution. In other words,
the `mode' may represent the kind of `texture' in a certain region
of the data, but two different regions with the same `texture'
should have independent realizations from the texture-generating
source. Our notion of a piecewise stationary process almost
coincides with that developed in \cite{Shamir}. The main difference
is that we allow an arbitrary distribution for the process
$\mathbf{A}$.

Now, we define $\mathcal{P}\{m_n\}$ to be  the class of all process
distributions that can be constructed as in \eq{PX def} for some
$M$, some collection
$P_{\mathbf{X}}^{(1)},\cdots,P_{\mathbf{X}}^{(M)}$ of stationary
processes, and some switching process $\mathbf{A}$  whose number of
shifts satisfies
\begin{equation}\label{eq: switching bound on growth rate}
r (A^n) \leq m_n \ \ \ \ a.s. \ \ \forall n.
\end{equation}
In words, a process $\mathbf{X}$ belongs to\footnote{The phrase
``the process $\mathbf{X}$ belongs to $\mathcal{P}\{m_n\}$'' is
shorthand for ``the distribution of the process $\mathbf{X}$,
$P_{\mathbf{X}}$, belongs to $\mathcal{P}\{m_n\}$''.}
$\mathcal{P}\{m_n\}$ if and only if it can be formed by switching
between a finite collection of independent processes in which the
number of switches by time $n$ does not exceed $m_n$.

\subsection{Richness of $\mathcal{P}
\{m_n\}$} \label{stochastic subsection 2}

In this subsection, we examine how rich the class
$\mathcal{P}\{m_n\}$ is, in terms of the growth rate $m_n$ and the
existence of denoising schemes that are universal with respect to
$\mathcal{P}\{m_n\}$. First, given any distribution on a noiseless
$n$-tuple, $P_{X^n}$, we define
\begin{equation}\label{eq: denoisability of n-tuple}
    \mathbb{D} (P_{X^n} , \mathbf{\Pi}) \triangleq \min_{\hat{\mathbf{X}}^n \in
    \mathcal{D}_n}  E L_{\hat{\mathbf{X}}^n} (X^n, Z^n) ,
\end{equation}
where $\mathcal{D}_n$ is the class of \emph{all} $n$-block
denoisers. The expectation on the right-hand side of \eq{eq:
denoisability of n-tuple} assumes that $X^n$ is generated from
$P_{X^n}$ and that $Z^n$ is the output of the DMC, $\mathbf{\Pi}$,
whose input is $X^n$. Thus, $\mathbb{D} (P_{X^n} , \mathbf{\Pi})$ is
the optimum denoising performance (in the sense of expected
per-symbol loss) attainable when the source distribution $P_{X^n}$
is known.

What happens when the source distribution is unknown? Theorem 3 of
\cite{Dude}  established the fact that\footnote{When
$P_{\mathbf{X}}$ is stationary, the limit $\lim_{n \rightarrow
\infty} \mathbb{D} (P_{X^n} , \mathbf{\Pi}) \mug \mathbb{D}
(P_{\mathbf{X}} , \mathbf{\Pi}) $ was shown to exist in \cite{Dude}.
Thus, \eq{eq: dude asymptotic performance} was equivalently stated
as  $\lim_{n \rightarrow \infty}  E
L_{\hat{\mathbf{X}}^n_{\mathrm{DUDE}}} = \mathbb{D} (P_{\mathbf{X}}
, \mathbf{\Pi})$ in \cite[Theorem 3]{Dude}.}
\begin{equation}\label{eq: dude asymptotic performance}
\lim_{n \rightarrow \infty} \left[ E
L_{\hat{\mathbf{X}}^n_{\mathrm{DUDE}}} (X^n, Z^n) - \mathbb{D}
(P_{X^n} , \mathbf{\Pi}) \right] = 0 \ \ \ \ \  \textrm{for all
stationary }\  P_{\mathbf{X}} .
\end{equation}
Note that  our newly-defined class of processes, $\mathcal{P}
\{m_n\}$, is simply the class of all stationary processes if one
takes the sequence $m_n$ to be $m_n \equiv 0$ for all $n$. Thus,
assuming $m_n \equiv 0$, \eq{eq: dude asymptotic performance} is
equivalent to
\begin{equation}\label{eq: dude asymptotic performance in terms of pmn}
\lim_{n \rightarrow \infty} \left[ E
L_{\hat{\mathbf{X}}^n_{\mathrm{DUDE}}} (X^n, Z^n) - \mathbb{D}
(P_{X^n} , \mathbf{\Pi}) \right] = 0 \ \ \ \ \  \mbox{ for all}\ \
P_{\mathbf{X}} \in  \mathcal{P} \{m_n\}.
\end{equation}
At the other extreme, when $m_n = n$, $\mathcal{P} \{m_n\}$ consists
of all possible  (not necessarily stationary) processes. We can
observe this equivalence by having $M = |\mathcal{X}|$ processes
each be a constant process at a different symbol in $\mathcal{X}$,
and creating any process by switching to the appropriate symbol. In
this case, not only does \eq{eq: dude asymptotic performance in
terms of pmn} not hold for the DUDE, but clearly  \eq{eq: dude
asymptotic performance in terms of pmn} cannot hold under \emph{any}
sequence of denoisers. In other words, $\mathcal{P} \{m_n\}$ is far
too rich to allow for the existence of schemes that are universal
with respect to it.

It is obvious then that  $\mathcal{P} \{m_n\}$ is significantly
richer than the family of stationary processes  whenever $m_n$ grows
with $n$. It is of interest then to  identify the maximal growth
rate of $m_n$ that allows for the existence of schemes that are
universal with respect to $\mathcal{P} \{m_n\}$, and to find such a
universal scheme. In what follows, we offer a complete answer to
these questions. Specifically, we  show that if the growth rate of
$m_n$ allows for the existence of \emph{any} scheme which is
universal with respect to $\mathcal{P} \{m_n\}$, the S-DUDE is
universal, too.

\subsection{Universality of S-DUDE}\label{stochastic subsection 3}

Here, we state our stochastic setting result, which establishes the
universality of $(k,m)$-S-DUDE with respect to the class
$\mathcal{P}\{m_n\}$.
\begin{theorem} \label{th: stochastic setting universality of the
S-DUDE} Let $k= k_n$ and $m = m_n$ satisfy the growth rate condition
stipulated in  Theorem \ref{semi stochastic theorem}, in addition to
$\lim_{n \rightarrow \infty} k_n = \infty$. Then, the sequence of
denoisers $\{\hat{\mathbf{X}}_{\mathrm{univ}}^{n,k,m}\}$ defined in
Section \ref{sec sdude} satisfy
\begin{equation}\label{eq: s-dude universality}
\lim_{n \rightarrow \infty} \left[ E
L_{\hat{\mathbf{X}}^{n,k,m}_{\mathrm{univ}}} (X^n, Z^n) - \mathbb{D}
(P_{X^n} , \mathbf{\Pi}) \right] = 0 \ \ \ \ \mbox{ for all } \
P_{\mathbf{X}}\in\mathcal{P}\{m_n\}.
\end{equation}
\end{theorem}

\noindent\emph{Remark 1:} Recall that, as noted in Claim \ref{claim:
k m condition}, $m_n =o(n)$ together with appropriately slowly
growing $k=k_n$ is sufficient to guarantee the growth rate condition
stipulated in Theorem \ref{semi stochastic theorem}. Hence, by
Theorem \ref{th: stochastic setting universality of the S-DUDE},
$m=o(n)$ and the sufficiently slowly growing $k=k_n$ suffices for
\eq{eq: s-dude universality} to hold. Therefore, Theorem \ref{th:
stochastic setting universality of the S-DUDE} implies the existence
of schemes that are universal with respect to $\mathcal{P} \{m_n\}$
whenever $m_n$ increases sublinearly in $n$. Since, as discussed in
Subsection \ref{stochastic subsection 2}, no universal scheme exists
for $\mathcal{P}\{m_n\}$ when $m_n$ is linear in $n$, we conclude
that the sub-linearity of $m_n$ is the necessary and sufficient
condition for a universal scheme to exist with respect to
$\mathcal{P}\{m_n\}$. Moreover, Theorem \ref{th: stochastic setting
universality of the S-DUDE} establishes the strong sense of
optimality of the S-DUDE, as it shows that whenever
$\mathcal{P}\{m_n\}$ is universally ``competable'', the S-DUDE does
the job.  This fact is somewhat analogous to the situation in
\cite{Shamir}, where the optimality of the universal lossless coding
scheme presented therein for piecewise stationary sources was
established under the condition that $m=o(n)$.
\\
\noindent \emph{Remark 2:} A pointwise result
$$
\lim_{n\rightarrow\infty}\big[
L_{\hat{\mathbf{X}}^{n,k,m}_{\mathrm{univ}}} (X^n, Z^n) - \mathbb{D}
(P_{X^n} , \mathbf{\Pi}) \big] = 0\quad\textrm{a.s.}
$$
for all $P_{\mathbf{X}}\in\mathcal{P}\{m_n\}$, which is analogous to
\cite[Theorem 4]{Dude}, can also be derived. However, we omit such a
result here since the details required for stating it rigorously
would be convoluted, and its added value over the strong point-wise
result we have already established in the semi-stochastic setting
would be little.

\emph{Proof of Theorem \ref{th: stochastic setting universality of
the S-DUDE}:} See Appendix \ref{proof of th: stochastic setting
universality of the S-DUDE}. $\quad\blacksquare$



\section{Algorithm and Complexity}\label{sec algorithm}
\subsection{An Efficient Implementation of S-DUDE}
 In the preceding two sections, we gave strong asymptotic
performance guarantees for the new class of schemes, the S-DUDE.
However, the question regarding the practical implementation of
\eq{Skhat def}, i.e., obtaining
$$
\hat{\Sb}_{k,m}=\arg\min_{\mathbf{S}\in\mathcal{S}_{k,m}^n}\tilde{L}_{\Sb}(z^n),
$$
for fixed $k$, $m$ and $n$ remains and, at first glance, may seem to
be a difficult combinatorial optimization problem. In this section,
we devise an efficient two-pass algorithm, which yields \eq{Skhat
def} and performs denoising with linear complexity in the sequence
length $n$. A recursion similar to that in the first pass of the
algorithm we present  appears also in the study of tracking the best
expert in on-line learning \cite{HerbsterWarmuth, BousquetWarmuth}.

From the definition of $\mathcal{S}_{k,m}^n$, (\ref{Skmn def}), we
can see that obtaining (\ref{Skhat def}) is equivalent to obtaining
the best combination of single-symbol denoisers with at most
$m(\cb)$ shifts that minimizes the cumulative estimated loss along
$\{t:t\in\mathcal{T}(\cb)\}$, for each $\cb\in\Cb_k$. Thus, our
problem breaks down to $| \Cb_k |$ independent problems, each being
a problem of competing with the best combination of single-symbol
schemes allowing $m$ switches.

 To describe an
algorithm that implements this parallelization efficiently, we first
define variables. For $(k,m)$-S-DUDE, let $I=m+1, J=N+1$, where
$N=|\mcS|=|\mcZ|^{|\mcXhat|}$. Then, a matrix
$M_t\in\mathbb{R}^{I\times J}$ is defined for $k+1\leq t\leq n-2k$,
where $M_t(i,j)$ for $1\leq i \leq I$ and $1\leq j\leq J-1$
represents the minimum (un-normalized) cumulative estimated loss of
the sequence of single-symbol denoisers along the time index
$\{\tau:\tau\leq t, \cb_{\tau}=\cb_t\}$, allowing at most $(i-1)$
shifts between single-symbol denoisers and applying $s_t=j$.
Moreover, $M_t(i,J)$, for $1\leq i \leq I$, is the symbol-by-symbol
denoiser that attains the minimum value of the $i$-th row of $M_t$,
i.e., $\arg\min_{1\leq j\leq J-1}M_t(i,j)$. A time pointer
$T\in\mathbb{R}^D$, where $D=|\Cb_k|=|\mcZ|^{2k}$, is defined to
store the closest time index that has the same context as current
time, during the first and second pass. That is,
\begin{eqnarray}
T(\mathbf{c}_t)\triangleq \left\{
\begin{array}{cc}
\max\{\tau:\tau<t, \mathbf{c}_{\tau}=\mathbf{c}_t\}, & \textrm{when first pass} \\
\min\{\tau:\tau>t, \mathbf{c}_{\tau}=\mathbf{c}_t\}, & \textrm{when second pass}\\
\end{array}\right\}\label{Tc def}
\end{eqnarray}
We also define $r\in\mathbb{R}^D$ and $q\in\mathbb{R}^D$ as
variables for storing the pointer enabling our scheme to follow the
best combination of single-symbol denoisers during the second pass.
Thus, the total memory size required is $O(mNn+|\mcZ|^{2k})=O(m n)$
(assuming that $k$ satisfies the growth rate stipulated in the
previous sections, which implies $|\mcZ|^{2k} = o(n)$).

Our two-pass algorithm has ingredients from both the DUDE and from
the forward-backward recursions of hidden Markov models
\cite{EphraimMerhav2002} and, in fact, the algorithm becomes
equivalent to DUDE when $m=0$. The first pass of the algorithm runs
forward from $t=k+1$ to $t=n-k$, and updates the elements of $M_t$
recursively. The recursions have a natural dynamic programming
structure. For $2 \leq i\leq I$, $1\leq j\leq J-1$, $M_t(i,j)$ is
determined by
\begin{eqnarray}
M_t(i,j)=\ell(z_t,j)+\min\Big\{M_{T(\cb_t)}(i,j),M_{T(\cb_t)}(i-1,M_{T(\cb_t)}(i-1,J))\Big\},\label{eq:Mt
recursion}
\end{eqnarray}
that is, adding the current loss to the best cumulative loss up to
$T(\cb_t)$ along $\{\tau:\tau<t, \cb_{\tau}=\cb_t\}$. When $i=1$,
the second term in the minimum of (\ref{eq:Mt recursion}) is not
defined, and $M_t(i,j)$ just becomes
$\ell(z_t,j)+M_{T(\cb_t)}(i,j)$. The validity of (\ref{eq:Mt
recursion}) can be verified by observing that there are two possible
cases in achieving $M_t(i,j)$: either the $(i-1)$-th shift to the
single-symbol denoiser $j$ occurred before $t$, or it occurred at
time $t$. We can see that the first term in the minimum of
(\ref{eq:Mt recursion}) corresponds to the former case; the second
term corresponds to the latter. Obviously, the minimum of these two
(where ties may be resolved arbitrarily), leads to the value of
$M_t(i,j)$ as in (\ref{eq:Mt recursion}).
%
%
%
%
%
%
%
%
%
%
%
%
%
%
%
%
%
%
%
%
%
%
After updating all $M_t$'s during the first pass, the second pass
runs backwards from $t=n-k$ to $t=k+1$, and extracts
$\hat{\Sb}_{k,m}$ from $\{M_t\}_{t=k+1}^{n-2k}$ by  following the
best shifting between single-symbol denoisers. The actual denoising
(i.e., assembling  the reconstruction sequence $\hat{X}^n$) is also
performed in that pass. The pointers $r(\cb_t)$ and $q(\cb_t)$ are
updated recursively, and they track the best shifting point and
combination of single-symbol denoisers, respectively, for each of
the subsequences associated with the various contexts. A succinct
description of the algorithm is provided in Algorithm \ref{alg1}.
The time complexity of the algorithm is readily seen to be $O(m n)$
as well.

\begin{algorithm}[H]
\caption{The $(k,m)$-Shifting Discrete Denoising Algorithm}
\label{alg1}
\begin{algorithmic}
\REQUIRE $M_t(i,j)\in\mathbb{R}^{I\times J}$, \quad$k+1\leq t \leq
n-2k, 1\leq i\leq I, 1\leq j \leq J$,\quad
$T\in\mathbb{R}^{D}$,$r\in\mathbb{R}^{D}, q\in\mathbb{R}^D$,
$L\in\mathbb{R}$ \ENSURE
$\hat{\Sb}_k=\{s_{k,t}(\cb_t,\cdot)\}_{t=k+1}^{n-2k}$ in \eq{Skhat
def} and the denoised output $\{\hat{x}_t\}_{t=k+1}^{n-k}$ \STATE
$\tau(\mathbf{c})\Leftarrow\phi$ for all $\mathbf{c}\in\mathbf{C}_k$

 \FOR{$t=k+1$ to $n-2k$} \IF{$T(\mathbf{c}_t)= \phi$} \STATE
$M_t(i,j)\Leftarrow\ell(z_t,j) \quad \textrm{for $1\leq i\leq
I$,\quad$1\leq j\leq J-1$}$ \STATE $M_t(i,J)\Leftarrow
\arg\min_{1\leq j\leq J-1}M_t(i,j) \quad\textrm{for $1\leq i\leq
I$}$ \ELSE \STATE $M_{T(\mathbf{c}_t)}^{*}(i,j) \Leftarrow \left\{
\begin{array}{ll}
M_{T(\mathbf{c}_t)}(i,j) & \textrm{for $i=1$,\quad$1\leq j \leq J-1$} \\
\min \big\{M_{T(\mathbf{c}_t)}(i,j),M_{T(\mathbf{c}_t)}(i-1,M_{T(\mathbf{c}_t)}(i-1,J))\big\} & \textrm{for $2\leq i\leq I$,\quad$1\leq j \leq J-1$}\\
\end{array}\right\}$
\STATE $M_t(i,j) \Leftarrow M_{T(\mathbf{c}_t)}^{*}(i,j)+\ell(z_t,j)
\quad\textrm{for $1\leq i\leq I$,\quad$1\leq j\leq J-1$}$ \STATE
$M_t(i,J)\Leftarrow \arg\min_{1\leq j\leq J-1}M_t(i,j)
\quad\textrm{for $1\leq i\leq I$}$\ENDIF \STATE $T(\mathbf{c}_t)
\Leftarrow t$\ENDFOR \STATE $T(\mathbf{c})\Leftarrow\phi$ for all
$\mathbf{c}\in\mathbf{C}_k$

\FOR{$t=n-2k$ to $k+1$} \IF{$T(\mathbf{c}_t)=\phi$}
 \STATE $r(\cb_t)\Leftarrow I$,\quad $q(\cb_t)\Leftarrow M_t(r(\cb_t),J)$
\ELSE
 \STATE $L\Leftarrow M_{T(\cb_t)}(r(\cb_t),q(\cb_t))-\ell(z_t,q(\cb_t))$
 \IF{$L<M_t(r(\cb_t),q(\cb_t))$}
 \STATE $r(\cb_t)\Leftarrow r(\cb_t)-1$,\quad $q(\cb_t)\Leftarrow M_t(r(\cb_t),J)$
 \ENDIF
\ENDIF
 \STATE $T(\cb_t)\Leftarrow t$, \ \  $s_{k,t}(\cb_t,\cdot)\Leftarrow q(\cb_t)$ \STATE
$\hat{x}_t\Leftarrow s_{k,t}(\cb_t,z_t)$
 \ENDFOR
\end{algorithmic}
\end{algorithm}
\noindent

\subsection{Extending the S-DUDE to Multi-Dimensional Data}
As noted, our algorithm is essentially separately employing the same
algorithm to compete with the best shifting single-symbol denoisers,
 on each subsequence associated with each context. The overall
 algorithm is the result of parallelizing  the operations of the schemes
 for the different subsequences,
 which allows for a more efficient implementation than if these
 schemes were to be run completely independently of one another.
  This
characteristic of running the same algorithm in parallel along each
subsequence enables us to extend S-DUDE to the case of
multi-dimensional data: run the same algorithm along each
subsequence associated with each (this time multi-dimensional)
context. It should be noted, however, that the extension of the
S-DUDE to  the multidimensional case is not as straightforward as
the extension of the DUDE was, since, whereas the DUDE's output is
independent of the ordering of the data within each context, this
ordering may be very significant in its effect on the output and,
hence, the performance of S-DUDE. Therefore, the choice of  a scheme
for scanning the data and capturing its local spatial stationarity,
e.g., Peano-Hilbert scanning \cite{LZ86}, is an important ingredient
in extending S-DUDE to the denoising of  multi-dimensional data.
Findings from the recent study on universal scanning reported in
\cite{Cohen1, Cohen2} can be brought to bear on such an  extension.

\section{Experimentation}\label{sec exp}

In this section, we report some preliminary experimental results
obtained by applying S-DUDE to several kinds of noise-corrupted
data.

\subsection{Image denoising}
\begin{figure}[h]
\begin{center}
\includegraphics[width=0.2\textwidth]{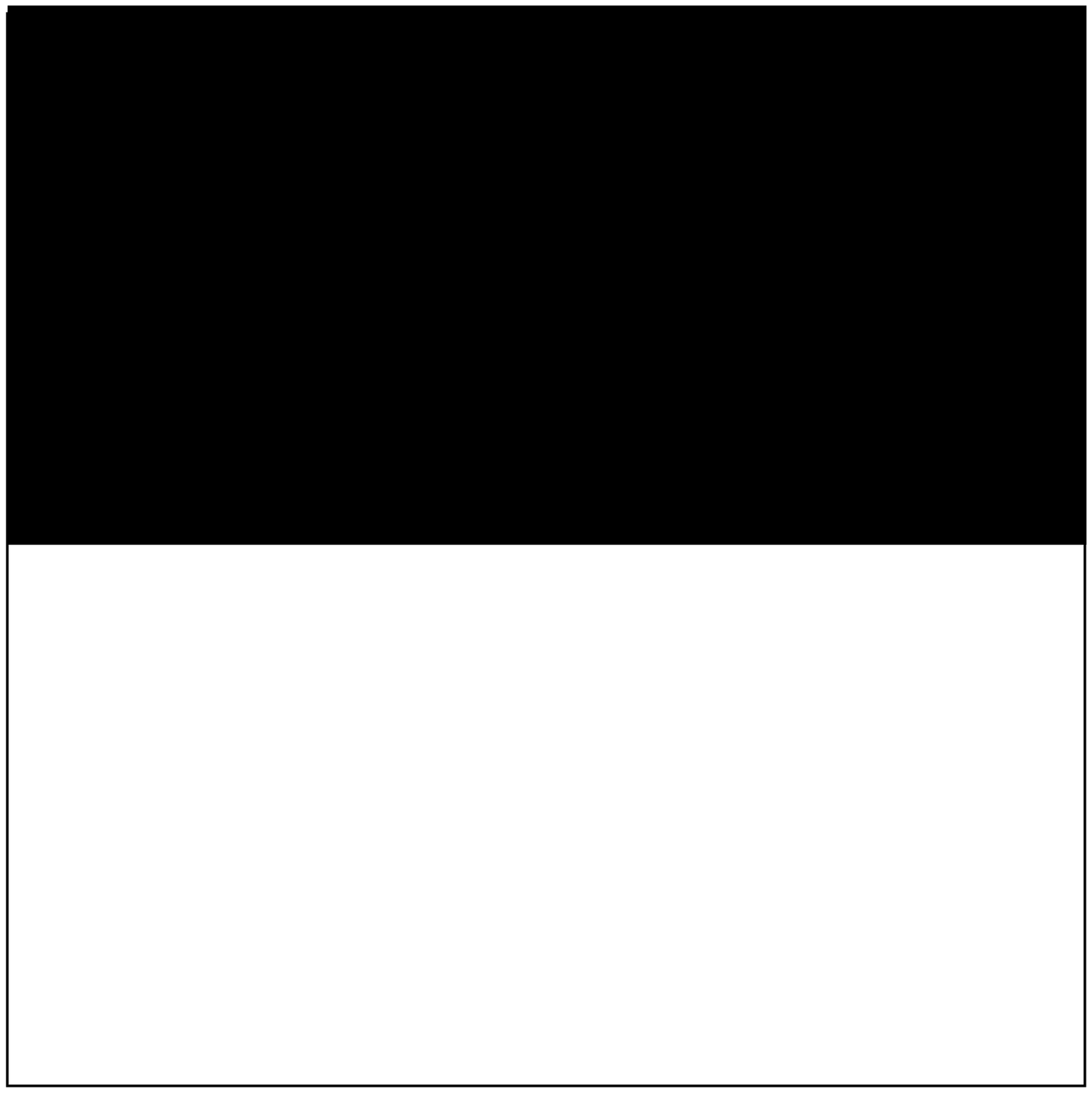}
\includegraphics[width=0.2\textwidth]{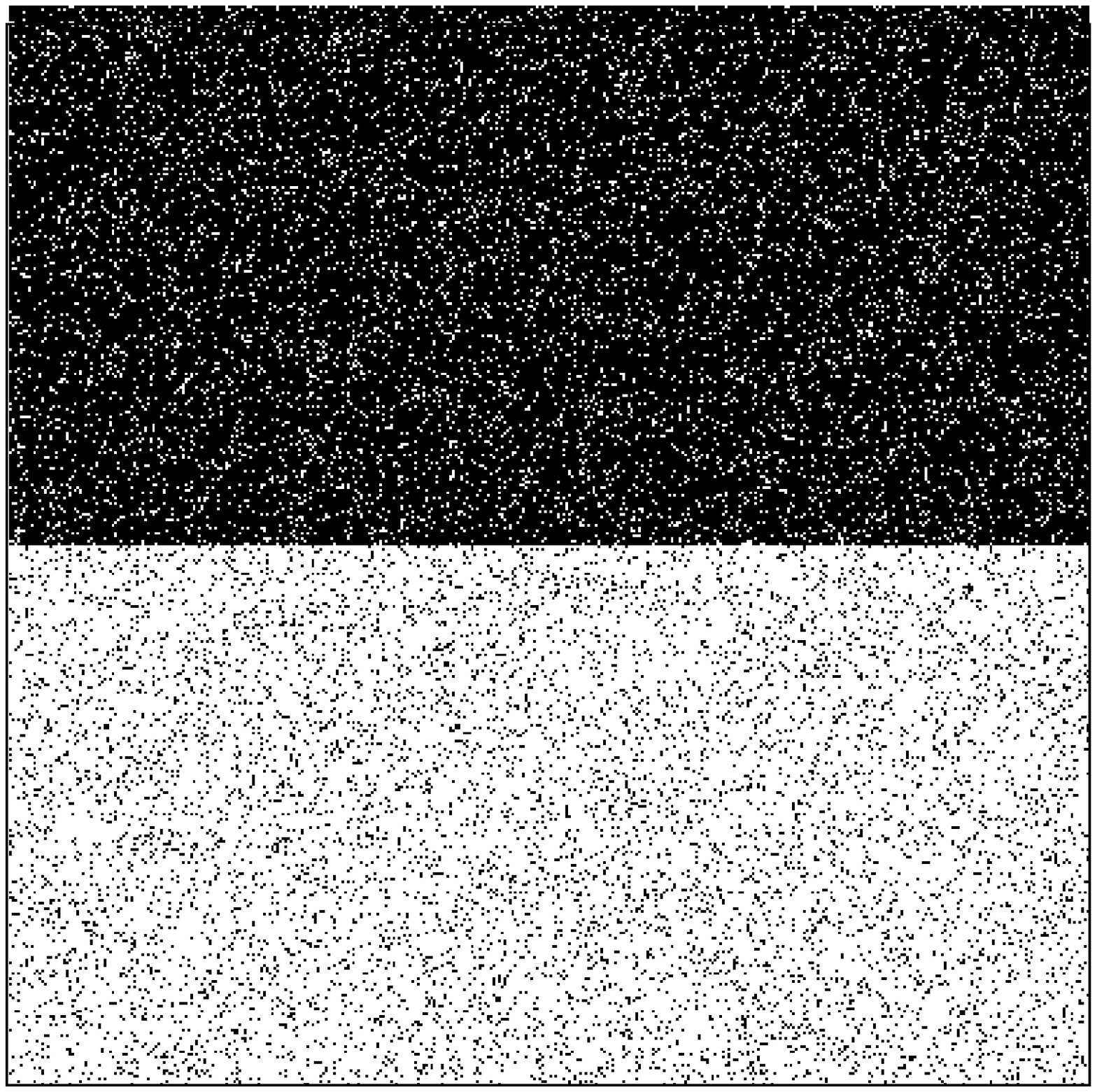}
\includegraphics[width=0.2\textwidth]{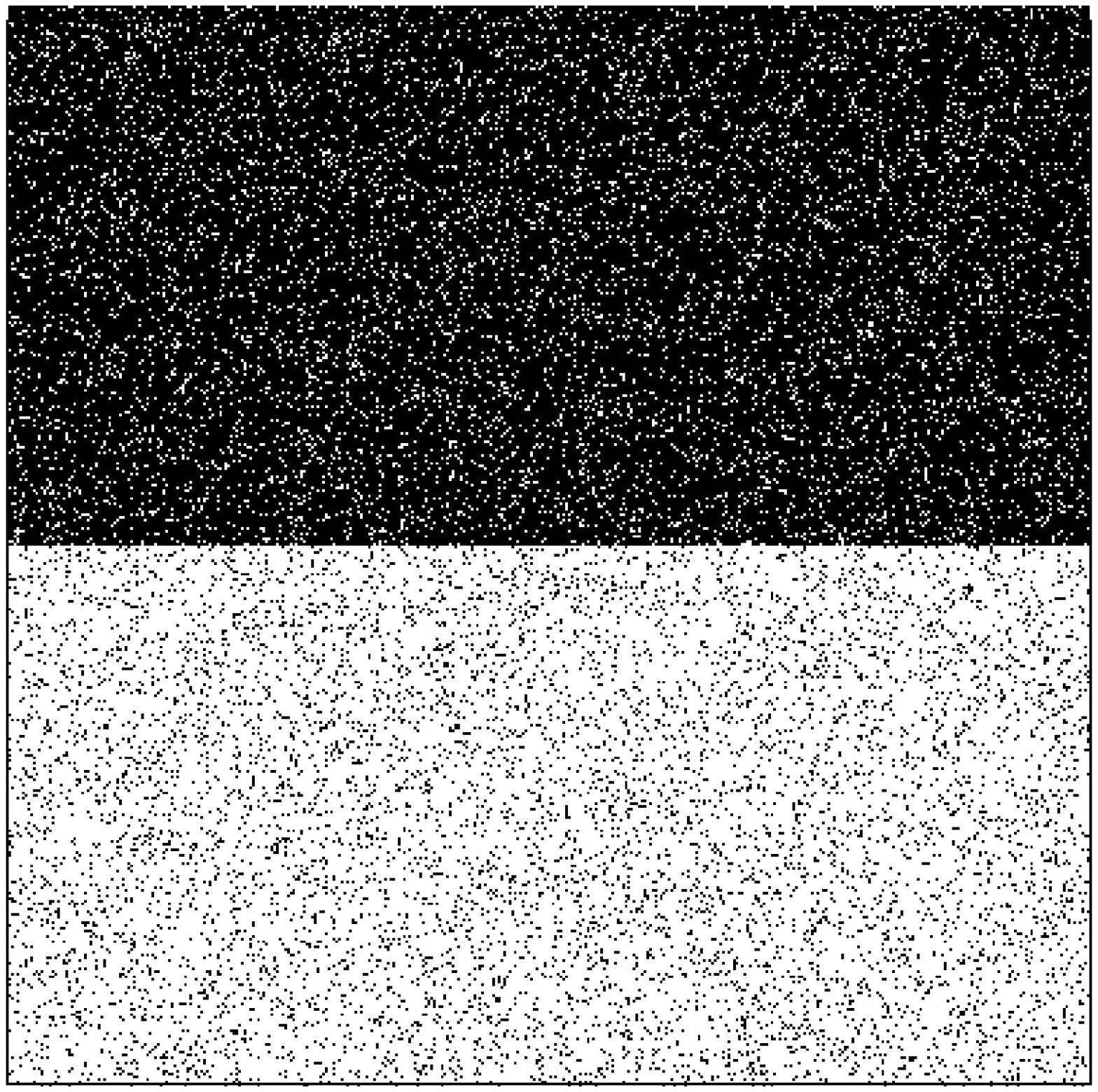}
\includegraphics[width=0.2\textwidth]{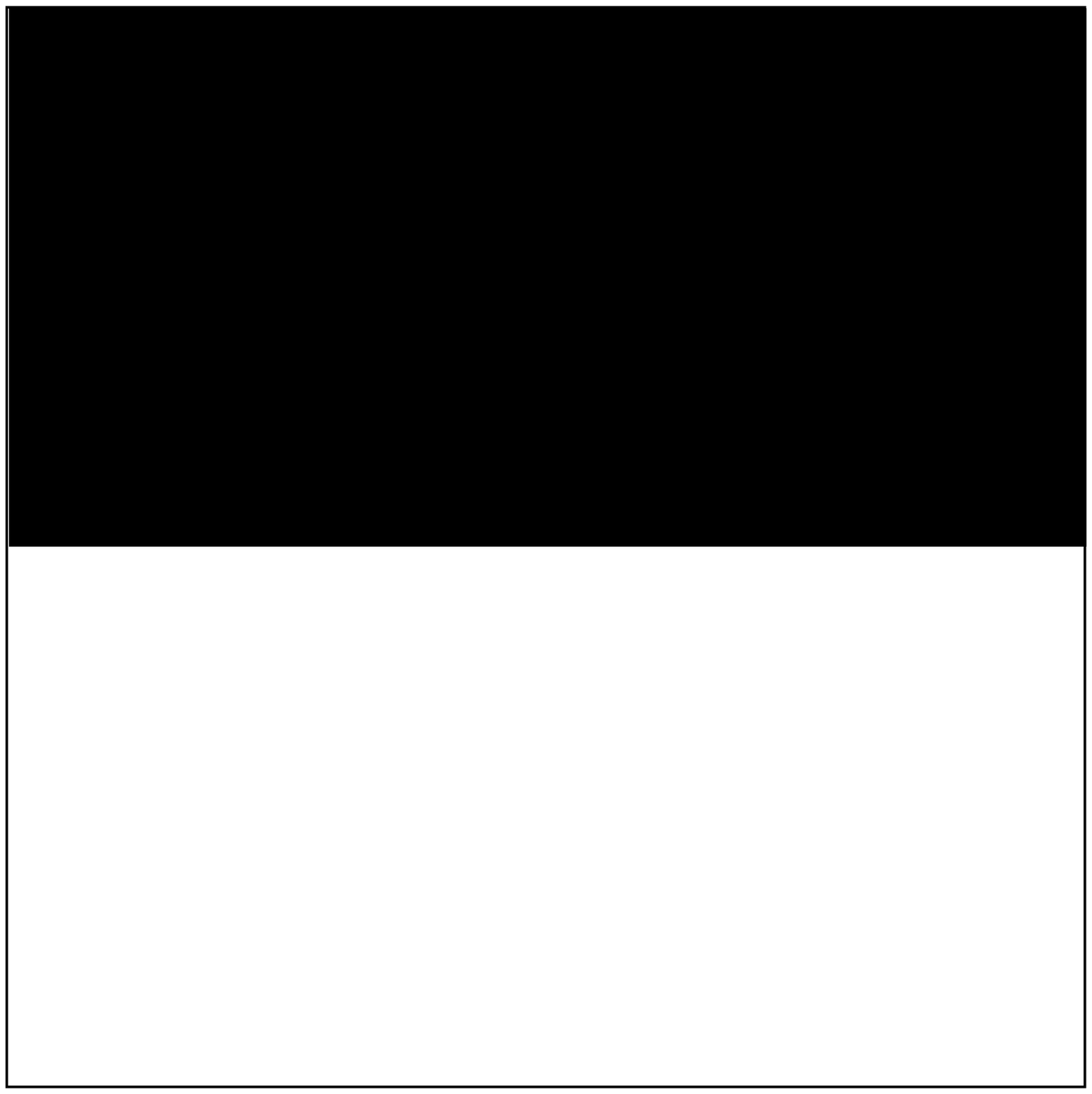}
\end{center}
\caption{$400\times 400$ binary images.}\label{binary image}
\end{figure}

In this subsection, we report  some experimental results of
denoising a binary image under the Hamming loss function. The first
and most simplistic experiment is with the  $400\times 400$
black-and-white binary image shown in Figure 1. The first figure is
the clean underlying image. The image is passed  through a binary
symmetric channel (BSC) with  crossover probability $\delta=0.1$, to
obtain the noisy image (second image in Figure \ref{binary image}).
Note that in this case, there are only four symbol-by-symbol
denoisers, namely, $\mcS=\{0,1,z,\bar{z}\}$, representing
always-say-$0$, always-say-$1$, say-what-you-see, and
flip-what-you-see, respectively. The third image in Figure
\ref{binary image} is the DUDE output with $k=0$, and the last image
is the output of our S-DUDE with $k=0,m=1$.
The DUDE with $k=0$ is competes with the best time-invariant
symbol-by-symbol denoiser which, in this case, is the
say-what-you-see denoiser, since the empirical distribution of the
clean image is $(0.5,0.5)$ and $\delta=0.1$. Thus, the DUDE output
is the same as the noisy image; hence, no denoising is performed.
However, it is clear that, for this image, the best compound action
of the symbol-by-symbol denoisers is always-say-$0$ for the first
half and then a shift to always-say-$1$ for the remainder. We can
see that our $(0,1)$-S-DUDE successfully captures this shift from
the noisy observations, and results in \emph{perfect} denoising with
zero bit errors.

\begin{figure}[h]
\begin{center}
\includegraphics[width=0.2\textwidth]{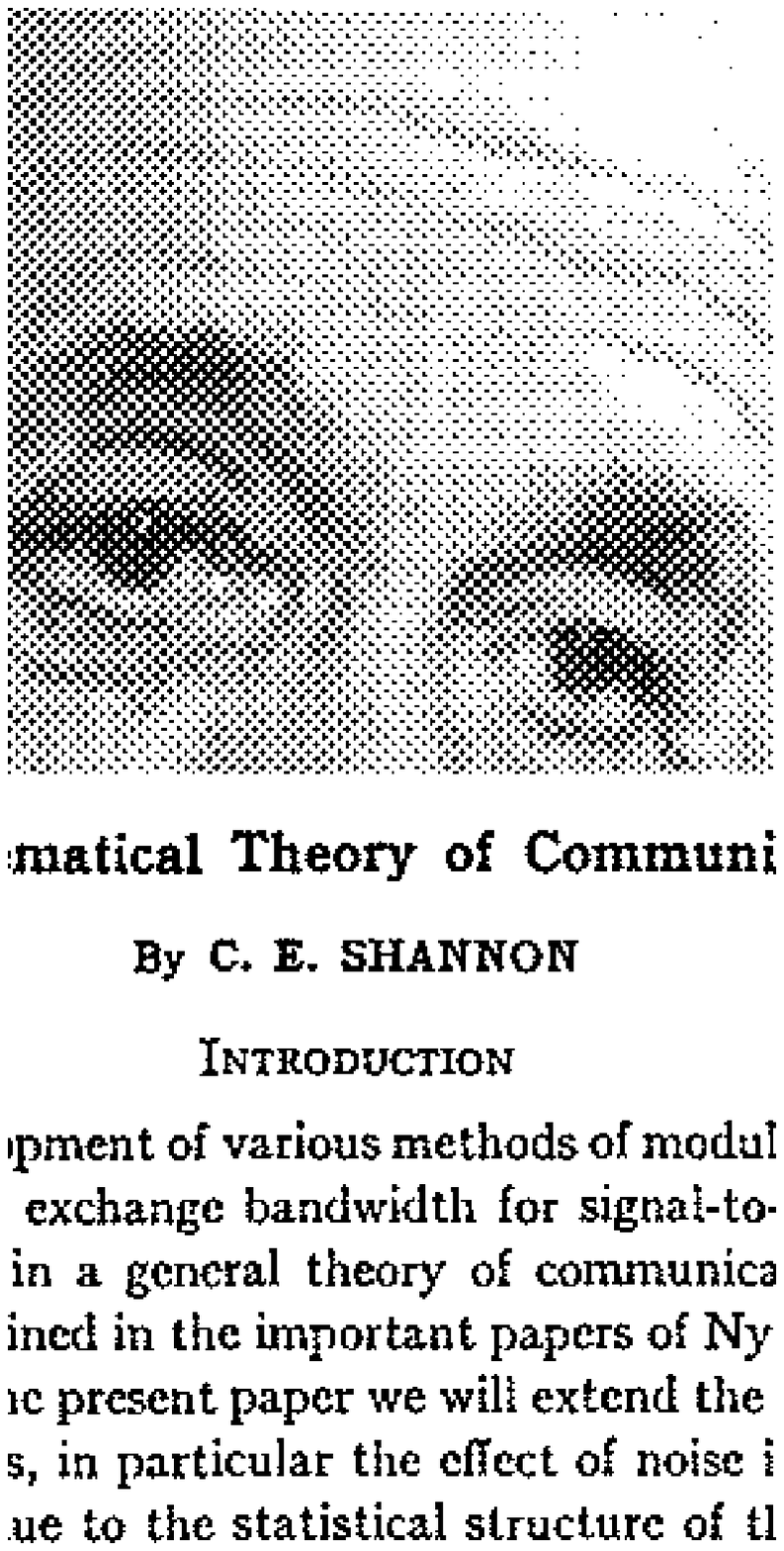}
\includegraphics[width=0.2\textwidth]{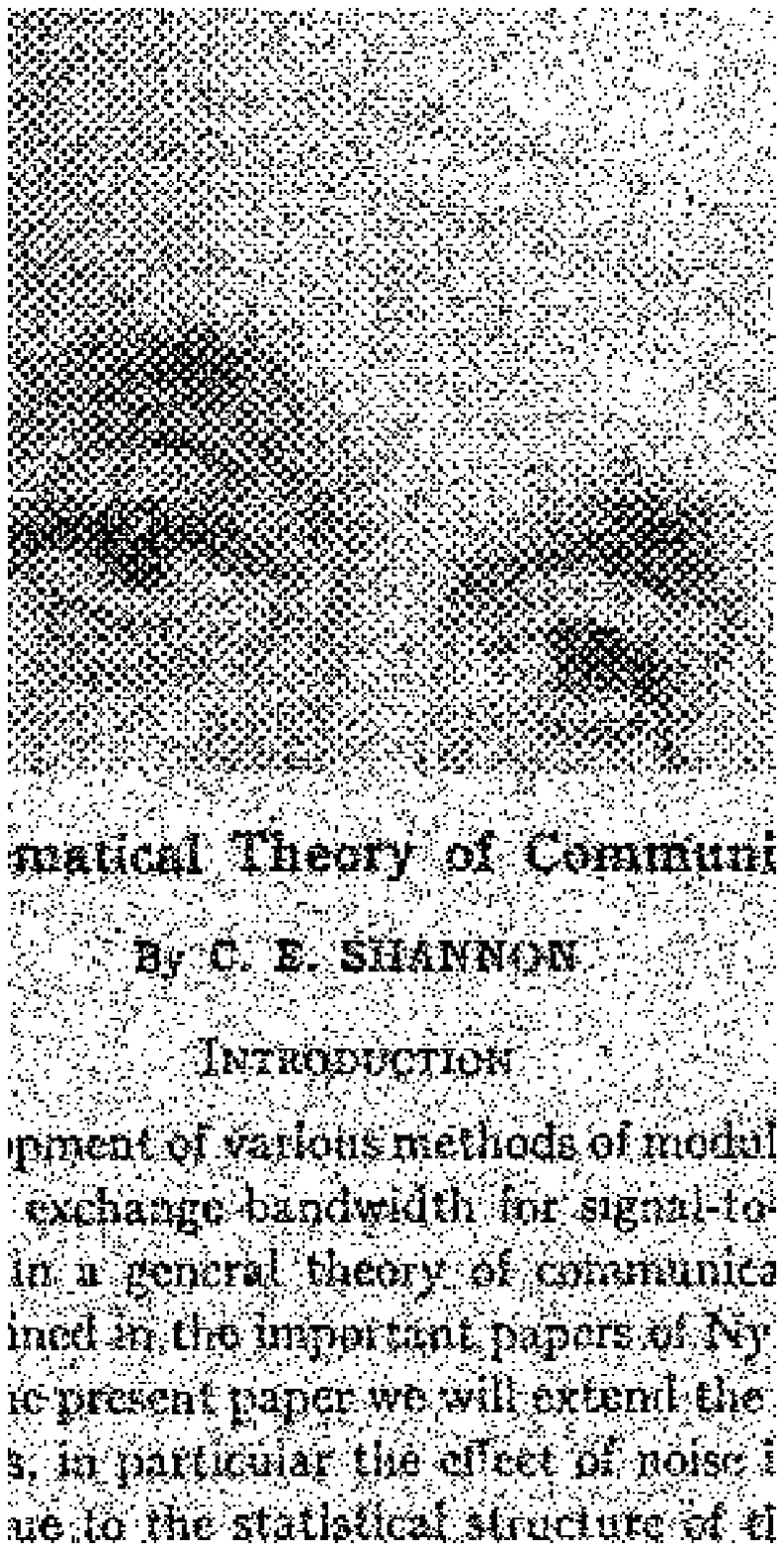}
\includegraphics[width=0.5\textwidth]{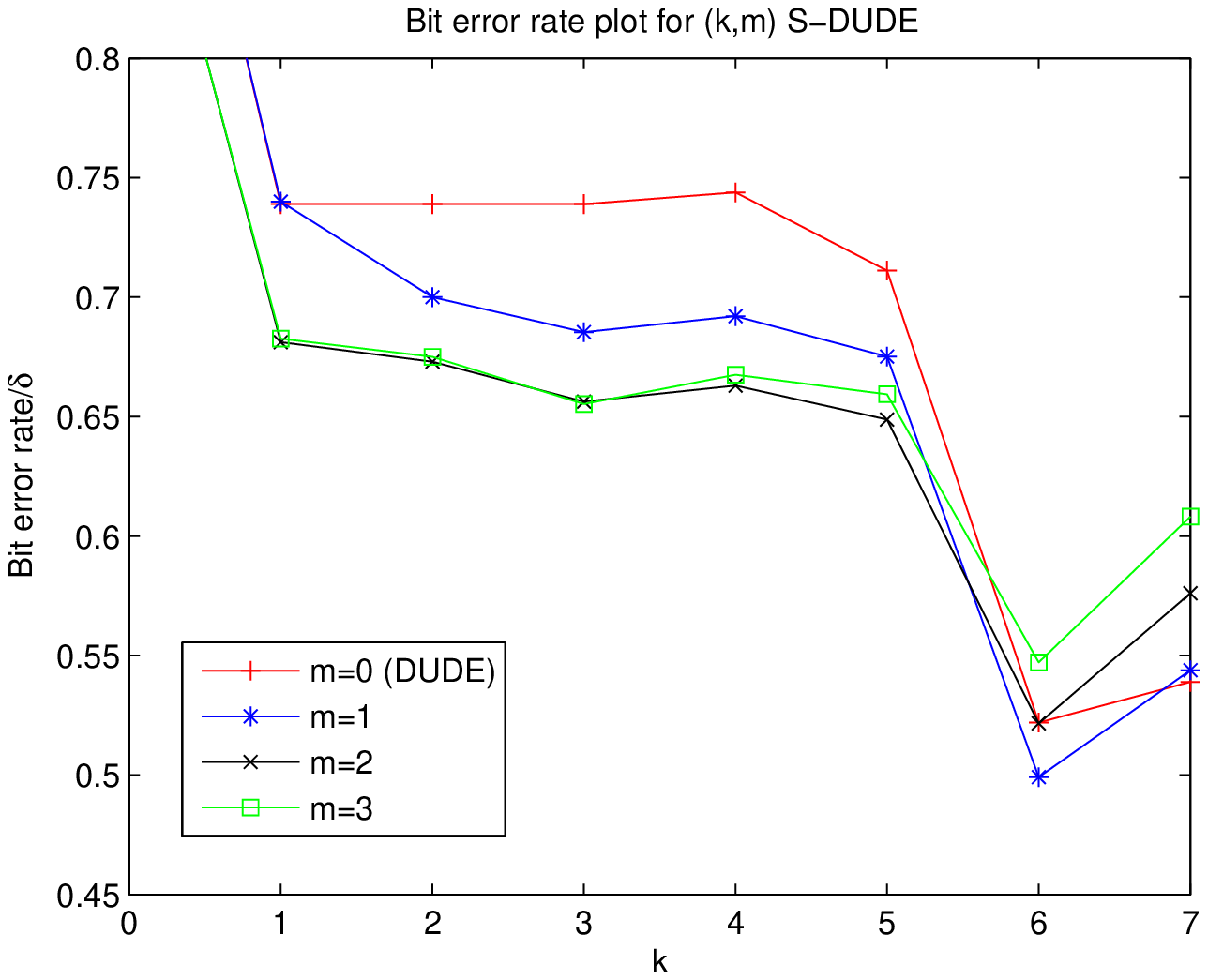}
\includegraphics[width=0.2\textwidth]{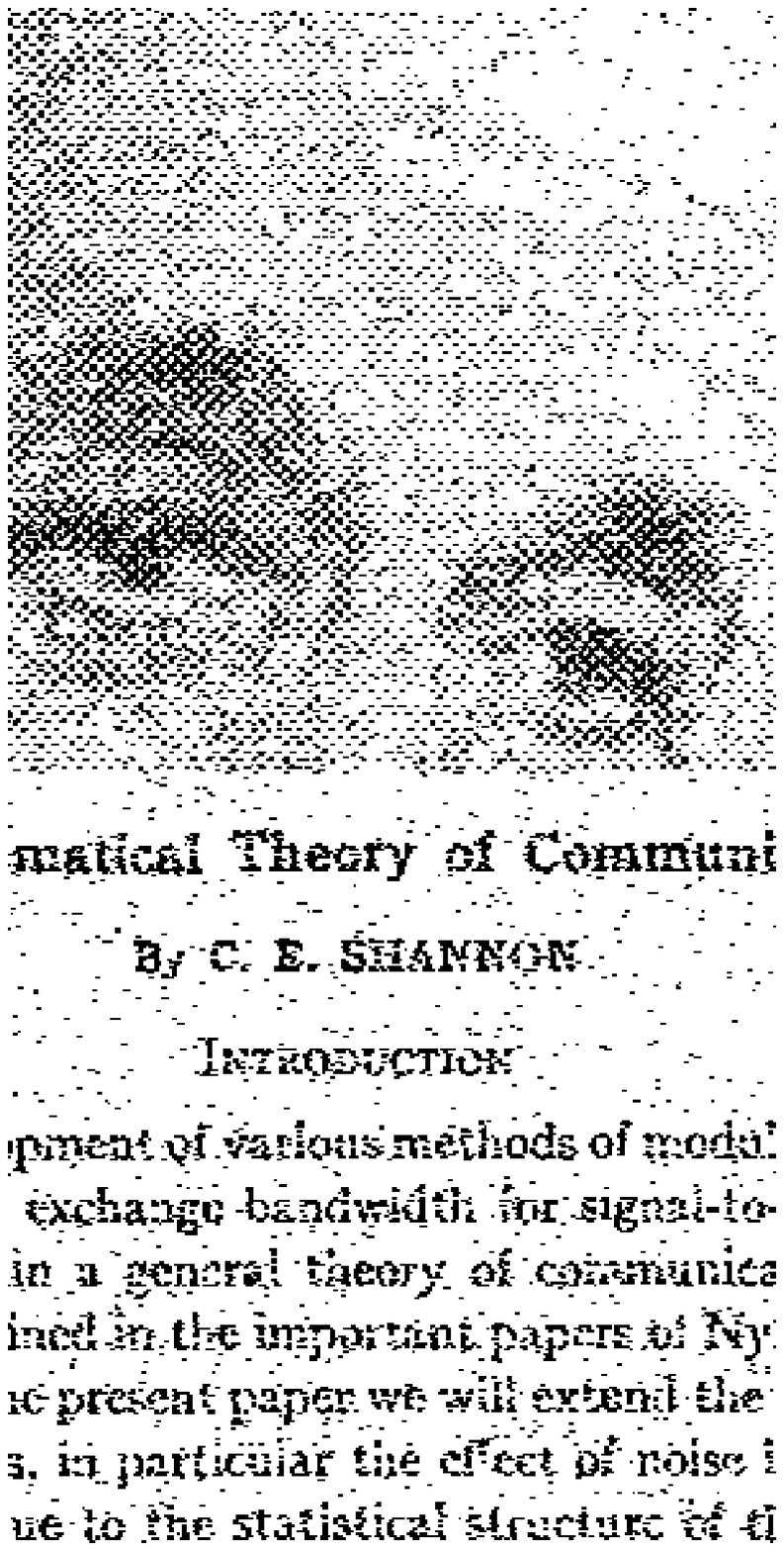}
\includegraphics[width=0.2\textwidth]{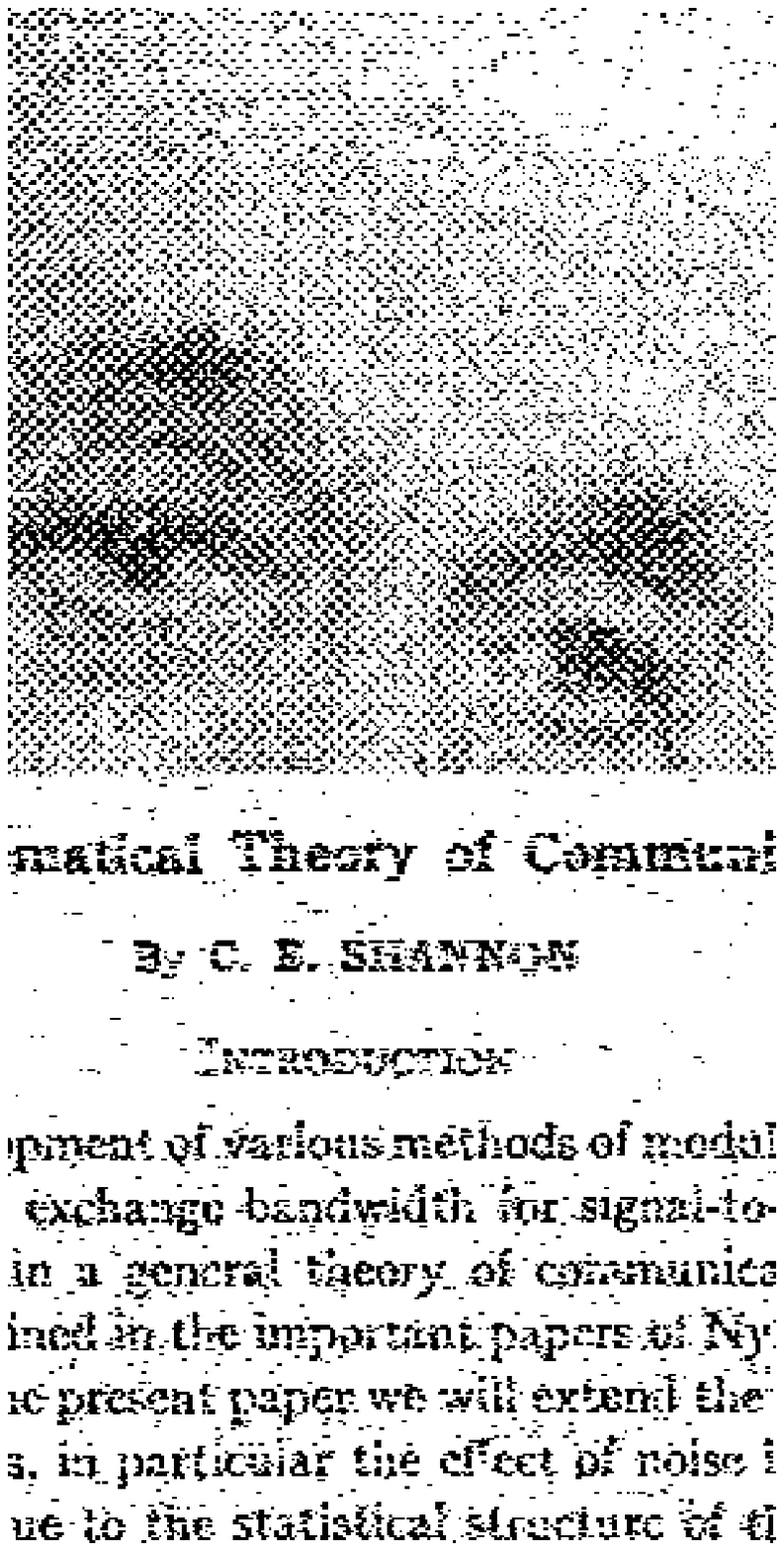}
\includegraphics[width=0.2\textwidth]{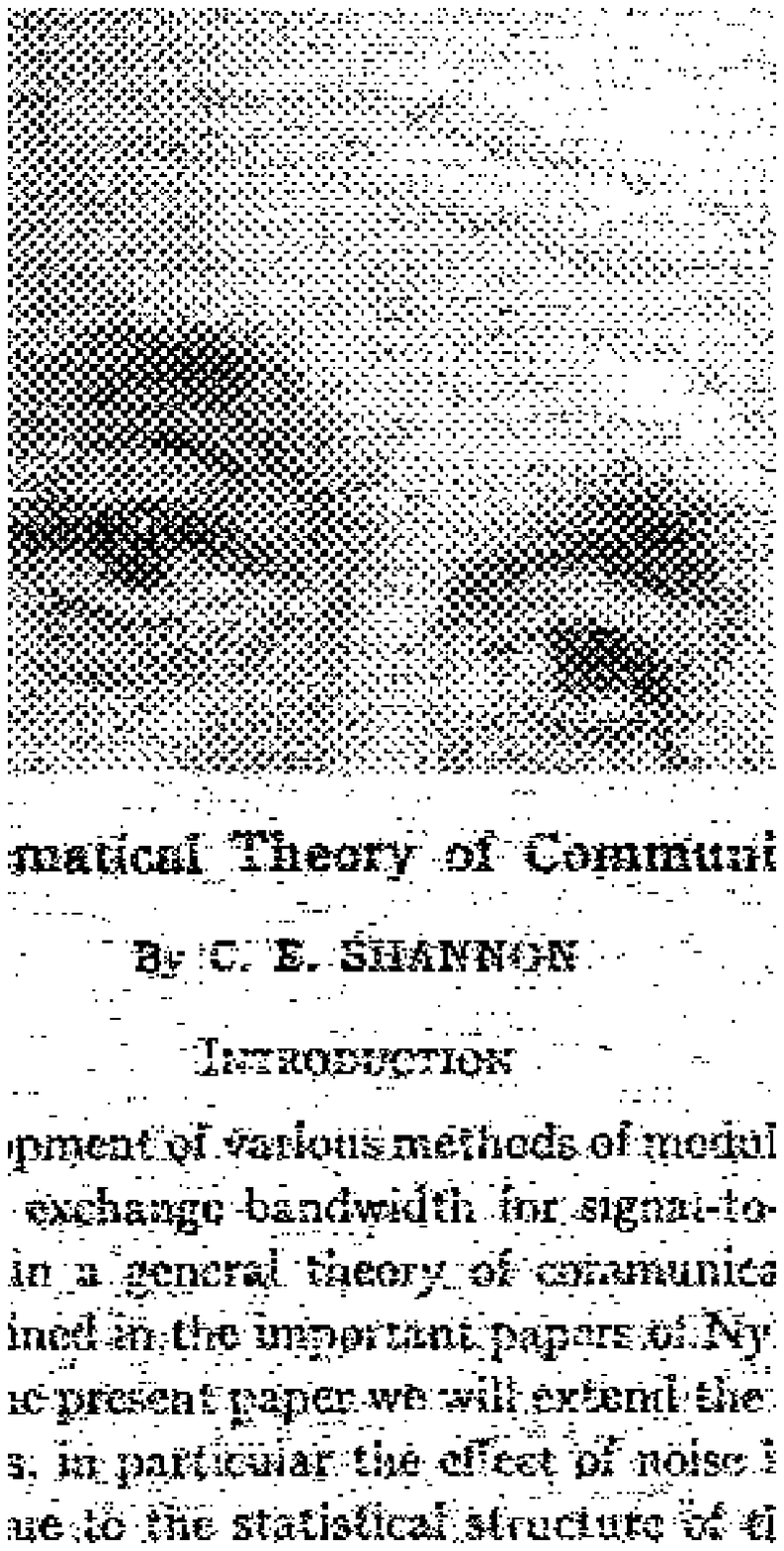}
\end{center}
\vspace{-.1in} \caption{Clean and noisy images, the bit error rate
plot for $(k,m)$-S-DUDE, and three denoised outputs for
$(k,m)=(4,0),(4,2),(6,1)$, respectively.}\label{figure:EinsteinShannon}
\end{figure}

Now, we move on to a more realistic example. The first image in
Figure \ref{figure:EinsteinShannon}, a concatenation of a half-toned
Einstein image ($300\times 300$) and scanned Shannon's 1948 paper
($300\times 300$), is the clean image. We pass the image through a
binary symmetric channel (BSC) with crossover probability
$\delta=0.1$, to obtain the second noisy image, which we raster scan
and employ the S-DUDE on the resulting one-dimensional sequence.
 Since the two concatenated images are of a very different nature, we
expect our S-DUDE to perform better than the DUDE, because it is
designed to adapt to the possibility of employing different schemes
in different regions of the data. The  plot shows the performance of
our $(k,m)$-S-DUDE with various values of $k$ and $m$. The
horizontal axis reflects $k$, and the vertical axis represents the
ratio of bit error per symbol (BER) to $\delta=0.1$. Each curve
represents the BER of schemes with different $m=0,1,2,3$. Note that
$m=0$ corresponds to the DUDE. We can see that S-DUDE with $m>0$
mostly dominates the DUDE, with an additional BER reduction of $\sim
11\%$, including when $k=6$, the best $k$ value for the DUDE. The
bottom three figures show the denoised images with
$(k,m)=(4,0),(4,2),(6,1)$, achieving BERs of $\delta\times
(0.744,0.6630,0.4991)$, respectively. Thus, in this example,
$(4,2)$-S-DUDE achieves an additional BER reduction of $11\%$ over
the DUDE with $k=4$, and the overall best performance is achieved by
$(6,1)$-S-DUDE. Given the nature of the image, which is a
concatenation of two completely different types of images, each
reasonably uniform in texture, it is not surprising to find that
 the S-DUDE with $m=1$ performs the best.

\subsection{State estimation for a switching binary hidden Markov process}
Here, we give a stochastic setting experiment. A switching binary
hidden Markov process in this example is defined as a binary
symmetric Markov chain observed through a BSC, where the transition
probabilities of the Markov chain switches over time. The goal of a
denoiser here is to estimate the underlying Markov chain based on
the noisy output.

\begin{figure}[h]
\begin{center}
\includegraphics[width=0.6\textwidth]{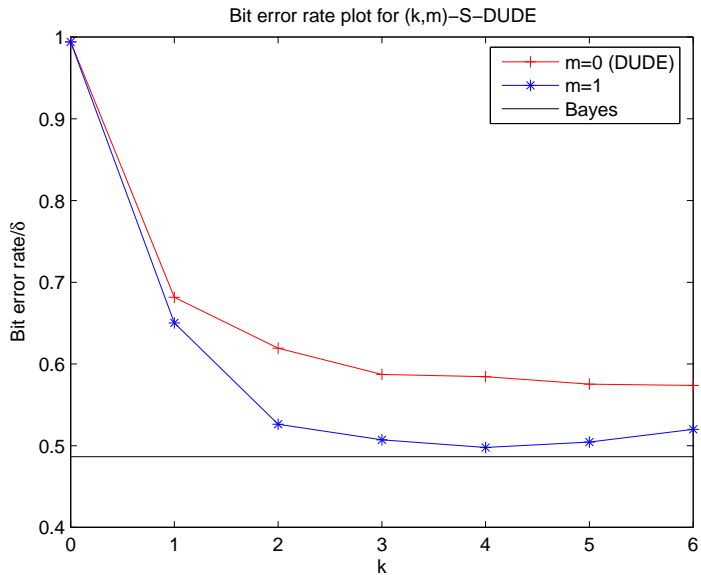}
\end{center}
\vspace{-.1in} \caption{BER for switching binary hidden Markov
process ($\delta=0.1,n=10^6$). The switch of the underlying binary
Markov chain occurs when $t=5\times10^5$, from the transition
probability $p=0.01$ to $p=0.2$.}\label{hmp result}
\end{figure}

In our example, we construct a simple switching binary hidden Markov
process of length $n=10^6$, in which the transition probability of
the underlying binary symmetric Markov source switches from $p=0.01$
to $p=0.2$ at the midpoint of the sequence, and the crossover
probability of BSC is $\delta=0.1$. Then, we estimate the state of
the underlying Markov chain based on the BSC output. The goodness of
the estimation is again measured by the Hamming loss, i.e., the
fraction of errors made.  Slightly better than the optimal Bayesian
distribution-dependent performance for this case can be obtained by
employing the forward-backward recursion scheme, incorporating the
varying transition probabilities with the help of a genie that knows
the exact location of the change in the process distribution. Figure
\ref{hmp result} plots the BER of $(k,m)$-S-DUDE with various $k$
and $m$, compared to the genie-aided Bayes optimal BER. The
horizontal axis represents $k$, and the two curves refer to $m=0$
(DUDE) and $m=1$. The vertical axis is the ratio of BER to
$\delta=0.1$.

We can observe that the optimal Bayesian BER is (lower bounded by)
$0.4865\times\delta$. The best performance of the DUDE was achieved
when $k=6$ with a BER of $0.5738\times\delta$, which is far above
(18\% more than)  the optimal BER. It is clear that, despite the
size of the data, the DUDE fails to converge to the optimum, as it
is confined to be employing the same  sliding-window scheme
throughout the whole data. However, we can see that the
$(4,1)$-S-DUDE achieves a BER of $.4979\times\delta$, which is
within $2.3\%$ of the optimal BER. This example shows that our
S-DUDE is competent in attaining the optimum performance for a class
richer than that of the stationary processes. Specifically, it
attains the optimum performance for piecewise stationary processes,
on which the DUDE generally fails.

\section{Conclusion and Some Future Directions}\label{sec conclusion}
Inspired by the DUDE algorithm, we have developed a generalization
that accommodates switching between sliding window rules. We have
shown a strong semi-stochastic setting result for our new scheme in
competing with shifting $k$-th order denoisers. This result implies
a stochastic setting result as well, asserting that the S-DUDE
asymptotically attains the optimal distribution-dependent
performance for the case in which the underlying data is piecewise
stationary. We also described an efficient low-complexity
implementation of the algorithm, and presented some simple
experiments that demonstrate the potential benefits of employing
S-DUDE in practice.

There are several future research directions related to this work.
The S-DUDE can be thought of as a generalization of the DUDE, with
the introduction of a new component captured by the non-negative
integer parameter $m$. Many previous extensions of the DUDE, such as
the settings of channel with memory\cite{zhang}, channel uncertainty
\cite{george}, applications to  channel decoding\cite{channel
decoding}, discrete-input, continuous-output data\cite{dembo},
denoising of analog data\cite{Kamakshi}, and decoding in the
Wyner-Ziv problem\cite{shirin}, may stand to benefit from a revision
that would incorporate the viewpoint of switching between
time-invariant schemes. Particularly, extending S-DUDE to the case
where the the data are analog as in \cite{Kamakshi} will be
non-trivial and interesting  from both a theoretical and a practical
viewpoint. In addition, as mentioned in Section \ref{sec algorithm},
an extension of the S-DUDE to the case of multi-dimensional data is
not as straightforward as the extension of the DUDE was. Such an
extension should prove interesting and practically important.
Finally, it would be useful to devise guidelines, in the spirit of
those in \cite{yuverdu, OWW}, for the choice of $k$ and $m$ based on
$n$ and the noisy observation sequence $z^n$.

\section*{Acknowledgments}
The first author is grateful to Professor Manfred Warmuth for
introducing him to a substantial amount of related work on the
expert tracking problems in online learning.

\section{Appendix}\label{appendix}
\subsection{Proof of Lemma \ref{lem from martingale}}\label{proof of
lem from martingale} We first establish the fact that for all
$x^n\in\mcX^{n}$, and for fixed $\mathbf{S}\in\mcS_0^n$,
$$
\Big\{n\big[L_{\hat{\mathbf{X}}^{n,\mathbf{S}}}(x^n,Z^n)-\tL_{\mathbf{S}}(Z^n)\big]\Big\}_{n\geq1}
$$
is a $\{Z_n\}$-martingale. This is not hard to see by following:
\begin{eqnarray}
&
&E\Big(n[L_{\hat{\mathbf{X}}^{n,\mathbf{S}}}(x^n,Z^n)-\tL_{\Sb}(Z^n)]\big|Z^{n-1}\Big)\nonumber\\
&=&E\Big(\sum_{t=1}^n\Lambda(x_t,s_t(Z_t))-\sum_{t=1}^n\ell(Z_t,s_t)\big|Z^{n-1}\Big)\nonumber\\
&=&(n-1)[L_{\hat{\mathbf{X}}^{n-1,\mathbf{S}}}(x^{n-1},Z^{n-1})-\tL_{\Sb}(Z^{n-1})]+E\Big(\Lambda(x_n,s_n(Z_n))-\ell(Z_n,s_n)\big|Z^{n-1}\Big)\nonumber\\
&=&(n-1)[L_{\hat{\mathbf{X}}^{n-1,\mathbf{S}}}(x^{n-1},Z^{n-1})-\tL_{\Sb}(Z^{n-1})],\label{app1}
\end{eqnarray}
where \eq{app1} follows from the fact that $Z_n$ is independent of
$Z^{n-1}$, and $E\Lambda(x_n,s_n(Z_n))=E\ell(Z_n,s_n)$. Therefore,
$L_{\Sb}(x^n,Z^n)-\tL_{\Sb}(Z^n)$ is a normalized sum of bounded
martingale differences; therefore the inequalities (\ref{eq:lemma 1
eq1}) and (\ref{eq:lemma1 eq2}) follow directly from the
Hoeffding-Azuma inequality \cite[Lemma A.7]{CBL06}.
$\quad\blacksquare$
\subsection{Proof of Theorem \ref{concentration thm}}\label{proof of
concentration thm} Consider following chain of inequalities:
\begin{eqnarray}
& &P\Big(L_{\hat{\Xb}^{n,\Shat}}(x^n,Z^n)-D_{0,m}(x^n,Z^n)>\epsilon\Big)\nonumber\\
&=&P\Big(\max_{\Sb\in\Snm}\big\{L_{\hat{\Xb}^{n,\Shat}}(x^n,Z^n)-L_{\hat{\Xb}^{n,\Sb}}(x^n,Z^n)\big\}>\epsilon\Big)\nonumber\\
&\leq&\sum_{\Sb\in\Snm}P\Big(L_{\hat{\Xb}^{n,\Shat}}(x^n,Z^n)-L_{\hat{\Xb}^{n,\Sb}}(x^n,Z^n)>\epsilon\Big)\label{concen eq2}\\
&\leq&\underbrace{\sum_{\Sb\in\Snm}P\Big(L_{\hat{\Xb}^{n,\Shat}}(x^n,Z^n)-\tL_{\Shat}(Z^n)>\epsilon/2\Big)}_{\mathrm{(i)}}+\underbrace{\sum_{\Sb\in\Snm}P\Big(\tL_{\Shat}(Z^n)-L_{\hat{\Xb}^{n,\Sb}}(x^n,Z^n)>\epsilon/2\Big)}_{\mathrm{(ii)}},\label{concen
eq3}
\end{eqnarray}
where \eq{concen eq2} follows from the union bound, and \eq{concen
eq3} follows from adding and subtracting $\tL_{\Shat}(Z^n)$, and the
union bound. For term (i) in (\ref{concen eq3}),
\begin{eqnarray}
\mathrm{(i)}&\leq&\sum_{\Sb\in\Snm}P\Big(\max_{\Sb\in\mcS_{0,m}^n}\big\{L_{\hat{\Xb}^{n,\Sb}}(x^n,Z^n)-\tL_{\Sb}(Z^n)\big\}>\epsilon/2\Big)\label{concen eq8}\\
            &\leq&\sum_{\Sb\in\Snm}\sum_{\Sb\in\Snm}\exp\Big(-n\frac{\epsilon^2}{2L_{\max}^2}\Big),\label{concen
eq9}
\end{eqnarray}
where (\ref{concen eq8}) follows from
$L_{\hat{\Xb}^{n,\Shat}}(x^n,Z^n)-\tL_{\Shat}(Z^n)\leq\max_{\Sb\in\mcS_{0,m}^n}\big\{L_{\hat{\Xb}^{n,\Sb}}(x^n,Z^n)-\tL_{\Sb}(Z^n)\big\}$,
and (\ref{concen eq9}) follows from the union bound and
(\ref{eq:lemma 1 eq1}). Similarly, for term (ii) in (\ref{concen
eq3}),
\begin{eqnarray}
\mathrm{(ii)}&\leq&\sum_{\Sb\in\Snm}P\Big(\tL_{\Sb}(Z^n)-L_{\hat{\Xb}^{n,\Sb}}(x^n,Z^n)>\epsilon/2\Big)\label{concen eq11}\\
             &\leq&\sum_{\Sb\in\Snm}\exp\Big(-n\frac{\epsilon^2}{2L_{\max}^2}\Big),\label{concen
eq12}
\end{eqnarray}
where (\ref{concen eq11}) follows from
$\tL_{\Shat}(Z^n)\leq\tL_{\Sb}(Z^n)$ a.s., and \eq{concen eq12}
follows from (\ref{eq:lemma1 eq2}). Therefore, continuing \eq{concen
eq3}, we obtain
\begin{eqnarray}
\eq{concen eq3}&\leq&2\sum_{\Sb\in\Snm}\sum_{\Sb\in\Snm}\exp\Big(-n\frac{\epsilon^2}{2L_{\max}^2}\Big)\nonumber\\
&=&2\Big[\sum_{k=0}^m {n-1 \choose k}N(N-1)^k\Big]^2\exp\Big(-n\frac{\epsilon^2}{2L_{\max}^2}\Big)\label{concen eq6}\\
&\leq&2\exp\left(-n\Big[\frac{\epsilon^2}{2L_{\max}^2}-2h\Big(\frac{m}{n}\Big)-\frac{2(m+1)\ln
N}{n}\Big]\right)\label{concen eq7},
\end{eqnarray}
where \eq{concen eq6} follows from $|\Snm|=\sum_{k=0}^m{n-1\choose
k}N(N-1)^k$, and \eq{concen eq7} follows from $|\Snm|\leq
N^{m+1}\exp\big(nh(\frac{m}{n})\big)$. Hence, the theorem is proved.
$\quad\blacksquare$
\subsection{Proof of Lemma \ref{kl concent}}\label{proof of
kl concent} We will prove (\ref{eq: lemma3 eq1}) since the proof of
(\ref{eq: lemma3 eq2}) is essentially identical. As in \cite{Dude},
define
$$
\mathcal{I}_{d}\triangleq\{t:k+1\leq t\leq n-k, t\equiv d\mod
(k+1)\},
$$
whose cardinality is denoted $n_{d}=\lfloor(n-d-k)/(k+1)\rfloor$.
Then, by denoting $\Cb_t=(Z_{t-k}^{t-1},Z_{t+1}^{t+k})$, we start
the chain of inequalities,
\begin{eqnarray}
&
&\mathrm{Pr}\Big(L_{\hat{\Xb}^{n,\Sb_k}}(x^{n-k}_{k+1},Z^n)-\tL_{\Sb_k}(Z^n)>\epsilon\Big)\nonumber\\
&\leq&\mathrm{Pr}\Bigg(\sum_{d=0}^k\sum_{\tau\in\mathcal{I}_d}\Big\{\Lambda\big(x_{\tau},s_{k,\tau}(\Cb_\tau,Z\tau)\big)-\ell\big(Z_\tau,s_{k,\tau}(\Cb_\tau,\cdot)\big)\Big\}>(n-2k)\epsilon\Bigg)\label{k
concen eq1}\\
&\leq&\sum_{d=0}^k\mathrm{Pr}\Bigg(\sum_{\tau\in\mathcal{I}_d}\Big\{\Lambda\big(x_{\tau},s_{k,\tau}(\Cb_\tau,Z\tau)\big)-\ell\big(Z_\tau,s_{k,\tau}(\Cb_\tau,\cdot)\big)\Big\}>(n-2k)\gamma_d\epsilon\Bigg)\label{k
concen eq2},
\end{eqnarray}
where \eq{k concen eq1} follows from the triangle inequality, \eq{k
concen eq2} follows from the union bound, and $\{\gamma_d\}$ is a
set of nonnegative constants (to be specified later) satisfying
$\sum_d\gamma_d=1$. In the sequel, for simplicity, we will denote
$\Lambda\big(x_{\tau},s_{k,\tau}(\Cb_\tau,Z\tau)\big)$ and
$\ell\big(Z_\tau,s_{k,\tau}(\Cb_\tau,\cdot)\big)$ in (\ref{concen
eq2}) as $\Lambda_\tau$ and $\ell_\tau$, respectively. Now, the
collection of random variables $Z(d)$ is defined to be
$$
Z(d)\triangleq\{Z_t:1\leq t\leq n, t\notin\mathcal{I}_d\},
$$
and $z(d)\in \mcZ^{n-n_d}$ denotes a particular realization of
$Z(d)$. Then, by conditioning, we have
\begin{eqnarray}
\eq{k concen
eq2}&\leq&\sum_{d=0}^k\sum_{z(d)\in\mcZ^{n-n_d}}\mathrm{Pr}(Z(d)=z(d))\mathrm{Pr}\Bigg(\sum_{\tau\in\mathcal{I}_d}\big\{\Lambda_\tau-\ell_\tau\big\}>(n-2k)\gamma_d\epsilon\Bigg|Z(d)=z(d)\Bigg),\label{k
concen eq3}
\end{eqnarray}
and let $P_d$ denote the conditional probability of \eq{k concen
eq3}. Now, conditioned on $Z(d)=z(d)$,
$\{Z_\tau\}_{\tau\in\mathcal{I}_d}$ are all independent, and the
summation in $P_d$ beomes
$$\sum_{\tau\in\mathcal{I}_d}\Big\{\Lambda\big(x_{\tau},s_{k,\tau}(\cb_\tau,Z\tau)\big)-\ell\big(Z_\tau,s_{k,\tau}(\cb_\tau,\cdot)\big)\Big\},$$
which is the sum of the absolute differences of the true and
estimated losses of the symbol-by-symbol denoisers
$s_{k,\tau}(\cb_\tau,\cdot)$ over $\tau\in\mathcal{I}_d$. Thus, we
can apply (\ref{eq:lemma 1 eq1}), and obtain
\begin{eqnarray}
P_d&=&\mathrm{Pr}\Bigg(\sum_{\tau\in\mathcal{I}_d}\big\{\Lambda_\tau-\ell_\tau\big\}>n_d\cdot\frac{(n-2k)\gamma_d\epsilon}{n_d}\Bigg|Z(d)=z(d)\Bigg)\nonumber\\
&\leq&\exp\Big(-\frac{2(n-2k)^2\gamma_d^2\epsilon^2}{L_{\max}^2n_d}\Big).
\end{eqnarray}
Following \cite{Dude}, we choose
$\gamma_d=\frac{\sqrt{n_d}}{\sum_{j}\sqrt{n_j}}$, and from the
Cauchy-Schwartz inequality and $\sum_d n_d=n-2k$, we arrive at
$$\frac{n_d}{\gamma_d^2}\leq(k+1)\sum_{d=0}^kn_d=(k+1)(n-2k),$$
and, hence,
\begin{eqnarray}
P_d\leq\exp\Big(-\frac{2(n-2k)\epsilon^2}{(k+1)L_{\max}^2}\Big).\label{Pd
ineq}
\end{eqnarray}
Therefore, plugging \eq{Pd ineq} into \eq{k concen eq3}, we finally
have
$$
\eq{k concen
eq3}\leq(k+1)\exp\Big(-\frac{2(n-2k)\epsilon^2}{(k+1)L_{\max}^2}\Big),
$$
which proves the lemma. $\quad\blacksquare$
\subsection{Proof of Theorem \ref{k concentration thm}}\label{proof of
k concentration thm}
The proof resembles that of Theorem
\ref{concentration thm}.
Consider
\begin{eqnarray}
&
&\mathrm{Pr}\Big(L_{\hat{\Xb}^{n,\hat{\Sb}_{k,m}}}(x^{n-k}_{k+1},Z^n)-D_{k,m}(x^n,Z^n)>\epsilon\Big)\nonumber\\
&=&P\Big(\max_{\Sb\in\Sknm}\big\{L_{\hat{\Xb}^{n,\Skhat}}(x^{n-k}_{k+1},Z^n)-L_{\hat{\Xb}^{n,\Sb}}(x^{n-k}_{k+1},Z^n)\big\}>\epsilon\Big)\nonumber\\
&\leq&\sum_{\Sb\in\Sknm}P\Big(L_{\hat{\Xb}^{n,\Skhat}}(x^{n-k}_{k+1},Z^n)-L_{\hat{\Xb}^{n,\Sb}}(x^{n-k}_{k+1},Z^n)>\epsilon\Big)\label{k concent eq2}\\
&\leq&\sum_{\Sb\in\Sknm}\Big\{P\Big(L_{\hat{\Xb}^{n,\Skhat}}(x^{n-k}_{k+1},Z^n)-\tL_{\Skhat}(Z^n)>\frac{\epsilon}{2}\Big)+P\Big(\tL_{\Skhat}(Z^n)-L_{\hat{\Xb}^{n,\Sb}}(x^{n-k}_{k+1},Z^n)>\frac{\epsilon}{2}\Big)\Big\}\label{k concent eq3}\\
&\leq&2(k+1)\sum_{\Sb\in\Sknm}\sum_{\Sb\in\Sknm}\exp\bigg(-\frac{(n-2k)\epsilon^2}{2(k+1)L_{\max}^2}\bigg)\label{k concent eq5}\\
&=&2(k+1)\bigg[\sum_{k=0}^{m(\cb)} {n(\cb)-1 \choose
k}N(N-1)^k\bigg]^{2|\Cb_k|}\exp\bigg(-\frac{(n-2k)\epsilon^2}{2(k+1)L_{\max}^2}\bigg),\label{k
concent eq6}
\end{eqnarray}
where \eq{k concent eq2}-\eq{k concent eq3} follow similarly as in
\eq{concen eq2}-\eq{concen eq3}; \eq{k concent eq5} follows from
arguments similar to \eq{concen eq8}, \eq{concen eq11}, and Lemma
\ref{kl concent} (which plays the role that Lemma \ref{lem from
martingale} played there); and \eq{k concent eq6} follows from
$|\mcS_{k,m}^n|=\big[(\sum_{k=0}^{m(\cb)} {n(\cb)-1 \choose
k}N(N-1)^k\big]^{|\Cb_k|}$. Now, for all $\cb\in\Cb_k$,
\begin{eqnarray}
\sum_{k=0}^{m(\cb)} {n(\cb)-1 \choose k}N(N-1)^k&\leq&
N^{m+1}\exp\Big(n(\cb)h\Big(\frac{m(\cb)}{n(\cb)}\Big)\Big)\nonumber\\
                                          &\leq&N^{m+1}\exp\Big((n-2k)h\Big(\frac{m(\cb)}{n-2k}\Big)\Big)\label{switch size1}\\
                                          &\leq&N^{m+1}\exp\Big((n-2k)h\Big(\frac{m}{n-2k}\Big)\Big),\label{switch
                                          size2}
\end{eqnarray}
where \eq{switch size1} is based on the fact that
$\exp(nh(\frac{m}{n}))$ is an increasing function in $n$, and
\eq{switch size2} follows from $m\leq\lfloor\frac{n-2k}{2}\rfloor$.
Therefore, together with $|\Cb_k|=|\mcZ|^{2k}$, we have
\begin{eqnarray}
\eq{k concent
eq6}&\leq&2(k+1)\exp\bigg(-(n-2k)\cdot\Big[\frac{\epsilon^2}{2(k+1)L_{\max}^2}-2|\mcZ|^{2k}\cdot\Big\{h\Big(\frac{m}{n-2k}\Big)+\frac{(m+1)\ln
N}{n-2k}\Big\}\Big]\bigg) ,
\end{eqnarray}
which proves the theorem.$\quad\blacksquare$
\subsection{Proof of Claim \ref{claim: k m condition}}\label{proof of claim: k m condition}
 For part
a), to show the necessity first, suppose $c_1\geq\frac{1}{2\log
|\mcZ|}$. Then, from $|\mcZ|^{2k}=n^{\frac{2k\log|\mcZ|}{\log n}}$,
we have $2|\mcZ|^{2k}\cdot\{h(\frac{m}{n-2k})+\frac{(m+1)\ln
N}{n-2k}\}=\Omega\big(n^{\frac{2k\log|\mcZ|}{\log
n}}(\frac{m}{n})^{1-\delta}\big)$, which will grow to infinity as
$n$ grows, even when $m$ is fixed. Therefore, the right-hand side of
(\ref{k concent part b}) is not summable. On the other hand,
$k=c_1\log n$ with $c_1<\frac{1}{2\log |\mathcal{Z}|}$ is readily
verified to suffice for the summability, provided that $m=m_n$ grows
at any sub-polynomial rate, i.e., grows more slowly than
$n^{\alpha}$ for any $\alpha>0$ (e.g., $c_2\log n$).

For part b), to show the necessity, suppose $m=\Theta(n)$. Then,
$h(\frac{m}{n-2k})+\frac{(m+1)\ln N}{n-2k}=\Theta(1)$, and, thus,
for sufficiently small $\epsilon$,
$\frac{\epsilon^2}{2(k+1)L_{\max}^2}-|\mcZ|^{2k}\cdot\big\{h\big(\frac{m}{n-2k}\big)+\frac{(m+1)\log
N}{n-2k}\big\}<0$ even for $k$ fixed. Therefore, the right-hand side
of (\ref{k concent part b}) is not summable. Hence, $m=o(n)$ is
necessary for the summability.  For sufficiency, suppose $m=m_n$ is
any rate, such that $\lim_{n\rightarrow\infty}\frac{m_n}{n}=0$.
Then,
\begin{eqnarray}
&
&\frac{\epsilon^2}{2(k+1)L_{\max}^2}-2|\mcZ|^{2k}\cdot\Big\{h\Big(\frac{m}{n-2k}\Big)+\frac{(m+1)\log
N}{n-2k}\Big\}\nonumber\\
&=&\frac{1}{k}\Big\{\frac{\epsilon^2}{2(1+\frac{1}{k}L_{\max}^2)}-2k|\mcZ|^{2k}\cdot
O\Big(\big(\frac{m_n}{n}\big)^{1-\delta}\Big)\Big\}.\label{th5
remark}
\end{eqnarray}
Thus, if $k$ grows sufficiently slowly that
$k|\mcZ|^{2k}=o\big((\frac{n}{m_n})^{1-\delta}\big)$, then \eq{th5
remark} becomes positive for sufficiently large $n$, and the
right-hand side of (\ref{k concent part b}) becomes summable.
 $\quad\blacksquare$
\subsection{Proof of Theorem \ref{semi stochastic theorem}}\label{proof of semi stochastic theorem}
First, denote the random variable $A_{k,m}^n\triangleq
L_{\hat{\Xb}^{n,k,m}_{\mathrm{univ}}}(x_{k+1}^{n-k},Z^n)-D_{k,m}(x^n,Z^n)$.
Then, for part a), we have
$$
L_{\hat{\Xb}^{n,k,m}_{\mathrm{univ}}}(x^n,Z^n)-D_{k,m}(x^n,Z^n)\leq
\frac{2k\Lambda_{\max}}{n}+A_{k,m}^n\quad\textrm{\emph{a.s.}}
$$
Since the maximal rate for $k$ is $c_1\log n$ as specified in Claim
\ref{claim: k m condition},
$\lim_{n\rightarrow\infty}\frac{2k\Lambda_{\max}}{n}=0$.
Furthermore, from the summability condition on $k$ and $m$, Theorem
\ref{k concentration thm}, and the Borel-Cantelli lemma, we get
$\lim_{n\rightarrow\infty}A_{k,m}^n=0$ with probability 1, which
proves  part a). To prove part b), note that, for any $\epsilon>0$,
\begin{eqnarray}
&
&E\big[L_{\hat{\Xb}^{n,k,m}_{\mathrm{univ}}}(x^n,Z^n)-D_{k,m}(x^n,Z^n)\big]\nonumber\\
&\leq&\frac{2k\Lambda_{\max}}{n}+E(A_{k,m}^n)\nonumber\\
&=&\frac{2k\Lambda_{\max}}{n}+E(A_{k,m}^n|A_{k,m}^n\leq\epsilon)Pr(A_{k,m}^n\leq\epsilon)+E(A_{k,m}^n|A_{k,m}^n>\epsilon)Pr(A_{k,m}^n>\epsilon)\nonumber\\
            &\leq&\frac{2k\Lambda_{\max}}{n}+\epsilon+\Lambda_{\max}\cdot Pr(A_{k,m}^n>\epsilon)\nonumber\\
            &\leq&\frac{2k\Lambda_{\max}}{n}+\epsilon+\Lambda_{\max}\cdot(\textrm{right-hand side of (\ref{k concent part b})}).\label{eq:thm4 expectation}
\end{eqnarray}
From the proof of Claim \ref{claim: k m condition}, the condition of
Theorem \ref{semi stochastic theorem} requires $k=k_n$ and $m=m_n$
to satisfy
$$\lim_{n\rightarrow\infty}k_n|\mcZ|^{2k_n}(\frac{m_n}{n})^{1-\delta}=0.$$
Therefore, if we set
$\epsilon^2=\Theta(k_n|\mcZ|^{2k_n}(\frac{m_n}{n})^{1-\delta})$ with
sufficiently large constant then, from (\ref{th5 remark}), we can
see that the right-hand side of (\ref{k concent part b}) will decay
almost exponentially, which is much faster than
$\Theta(k_n|\mcZ|^{2k_n}(\frac{m_n}{n})^{1-\delta})$. Hence, from
(\ref{eq:thm4 expectation}), we conclude that
$E(A_{k,m}^n)=O\left(\sqrt{k_n|\mcZ|^{2k_n}(\frac{m_n}{n})^{1-\delta}}\right)$,
which results in part b). $\quad\blacksquare$

\subsection{Proof of Theorem \ref{th: converse establishing necessity}}\label{proof of th: converse establishing necessity}
The fact that $m = \Theta (n)$ implies the existence of $\alpha >
0$, such that $m \geq n \alpha$ for all sufficiently large $n$. Let
$\mathbf{X}$ be the process formed by concatenating i.i.d.\ blocks
of length $\lceil 1/\alpha \rceil$, each block consisting  of the
same repeated symbol chosen uniformly from $\mathcal{X}$. The first
observation to note is that, for all $n$ large enough that $m \geq n
\alpha$,
\begin{equation}\label{eq: a.s. nailing of genie}
D_{0, m} (X^n, Z^n) = 0   \ \ \ \ a.s.
\end{equation}
This is because, by construction, $X^n$ is, with probability 1,
piecewise constant with constancy sub-blocks of length, at least,
$\lceil 1/\alpha \rceil$. Thus, a genie with access to $X^n$ can
choose a sequence of symbol-by-symbol schemes (in fact, ignoring the
noisy sequence), with less than $n \alpha$ (and, therefore, less
than $m$) switches, that perfectly recover $X^n$ (and, therefore, by
our assumption on the loss function, suffers zero loss). On the
other hand, the assumptions on the loss function and the channel
imply that, for the process $\mathbf{X}$ just constructed,
\begin{equation}\label{eq: even bayesian guy fails}
\limsup_{n\rightarrow\infty} \min_{\hat{\mathbf{X}}^n}
EL_{\hat{\mathbf{X}}^n}(X^n,Z^n)> 0,
\end{equation}
since even the Bayes-optimal scheme for this process incurs a
positive loss, with a positive probability, on each $\lceil 1/\alpha
\rceil$ super-symbol. Thus, we get
\begin{eqnarray}
 & & E \Big\{ \limsup_{n \rightarrow \infty} E \left[ L_{\hat{\mathbf{X}}^n}(X^n, Z^n) - D_{0, m} (X^n, Z^n)| X^n \right] \Big\} \label{eq: exp to be extracted1} \\
 &\geq& \limsup_{n \rightarrow \infty} E \left[ L_{\hat{\mathbf{X}}^n}  (X^n, Z^n) - D_{0, m} (X^n, Z^n)
    \right] \label{eq: exp to be extracted2}  \\
&=& \limsup_{n \rightarrow \infty} E  L_{\hat{\mathbf{X}}^n}  (X^n, Z^n)   \label{eq: exp to be extracted3} \\
&\geq&  \limsup_{n \rightarrow \infty} \min_{\hat{\mathbf{X}}^n}
EL_{\hat{\mathbf{X}}^n}(X^n,Z^n)  \nonumber \\
&>& 0, \label{eq: exp to be extracted4}
\end{eqnarray}
where (\ref{eq: exp to be extracted2}) follows from Fatou's lemma;
(\ref{eq: exp to be extracted3}) follows from \eq{eq: a.s. nailing
of genie}; and (\ref{eq: exp to be extracted4}) follows from \eq{eq:
even bayesian guy fails}. In particular, there must be one
particular individual sequence $\mathbf{x} \in \mathcal{X}^\infty$
for which the expression inside the curled brackets of \eq{eq: exp
to be extracted1} is positive, i.e.,
\begin{equation}\label{eq: positivity of curled brack exp for someone}
\limsup_{n \rightarrow \infty} E \left[ L_{\hat{\mathbf{X}}^n} (X^n,
Z^n) - D_{0, m} (X^n, Z^n)
   | X^n = x^n \right] > 0,
\end{equation}
which is equivalent to \eq{eq: limsup of exp diff is positive}.
$\quad\blacksquare$
\subsection{Proof of Theorem \ref{th: stochastic setting universality of the S-DUDE}}\label{proof of th: stochastic setting universality of the S-DUDE}
First, by adding and subtracting the same terms, we obtain
\begin{eqnarray}
& &EL_{\hat{\mathbf{X}}_{\mathrm{univ}}^{n,k,m}}(X^n,Z^n)-\mathbb{D}(P_{X^n},\mathbf{\Pi})\nonumber\\
&=&\underbrace{EL_{\hat{\mathbf{X}}_{\mathrm{univ}}^{n,k,m}}(X^n,Z^n)-\min_{\mathbf{S}\in\mcS_{k,m}^n}EL_{\hat{\mathbf{X}}^{n,\mathbf{S}}}(X^n,Z^n)}_{\mathrm{(i)}}+\underbrace{\min_{\mathbf{S}\in\mcS_{k,m}^n}EL_{\hat{\mathbf{X}}^{n,\mathbf{S}}}(X^n,Z^n)-\mathbb{D}(P_{X^n},\mathbf{\Pi})}_{\mathrm{(ii)}}.\label{eq:
proof of theorem 5. Adding and Subtracting}
\end{eqnarray}
We will consider term (i) and term (ii) separately. For term (i),
\begin{eqnarray}
\mathrm{(i)}&=&EL_{\hat{\mathbf{X}}_{\mathrm{univ}}^{n,k,m}}(X^n,Z^n)-\min_{\mathbf{S}\in\mcS_{k,m}^n}EL_{\hat{\mathbf{X}}^{n,\mathbf{S}}}(X^n,Z^n)\nonumber\\
   &\leq&\frac{2k\Lambda_{\max}}{n}+\frac{n-2k}{n}\cdot \Big[EL_{\hat{\mathbf{X}}_{\mathrm{univ}}^{n,k,m}}(X^{n-k}_{k+1},Z^n)-\min_{\mathbf{S}\in\mcS_{k,m}^n}EL_{\hat{\mathbf{X}}^{n,\mathbf{S}}}(X^{n-k}_{k+1},Z^n)\Big]\label{th5 eq1}\\
   &\leq&\frac{2k\Lambda_{\max}}{n}+\frac{n-2k}{n}\cdot E\Big[L_{\hat{\mathbf{X}}_{\mathrm{univ}}^{n,k,m}}(X^{n-k}_{k+1},Z^n)-\min_{\mathbf{S}\in\mcS_{k,m}^n}L_{\hat{\mathbf{X}}^{n,\mathbf{S}}}(X^{n-k}_{k+1},Z^n)\Big]\label{th5 eq2}\\
   &\leq&\frac{2k\Lambda_{\max}}{n}+E\Big[L_{\hat{\mathbf{X}}_{\mathrm{univ}}^{n,k,m}}(X^{n-k}_{k+1},Z^n)-D_{k,m}(X^n,Z^n)\Big]\label{th5
   eq3},
\end{eqnarray}
where \eq{th5 eq1} follows from upper bounding and omitting the
losses for time instances $t\leq k$ and $t>n-k$ in the first and
second terms of (i), respectively; \eq{th5 eq2} follows from
exchanging the minimum with the expectation, and \eq{th5 eq3}
follows from the
definition \eq{Dkmin} and $\frac{n-2k}{n}\leq1$. 

For term (ii), we bound the first term in (ii) as
\begin{eqnarray}
& &\min_{\mathbf{S}\in\mcS_{k,m}^n}EL_{\hat{\mathbf{X}}^{n,\mathbf{S}}}(X^n,Z^n)\nonumber\\
&\leq&\frac{2k(m+1)\Lambda_{\max}}{n}+\frac{1}{n}\min_{\mathbf{S}\in\mcS_{k,m}^n}E\bigg[E\Big[\sum_{i=1}^{r+1}\sum_{j=\tau_{i-1}+k+1}^{\tau_i-k}\Lambda(X_j,s_{k,j}(Z_{j-k}^{j+k}))\Big|A^n\Big]\bigg],\label{mineq1}
\end{eqnarray}
by upper bounding the losses with $\Lambda_{\max}$ on the boundary
of the shifting points. Now, let
$\mathbf{P}_{X_j|Z_i^l,A^n}\in\mathbb{R}^{|\mathcal{X}|}$ denote the
$|\mcX|$-dimensional probability vector whose $x$-th component is
$Pr(X_j=x|Z_i^l,A^n)$. Then, we can bound the second term in
\eq{mineq1} by the following chain of inequalities:
\begin{eqnarray}
&
&\frac{1}{n}\min_{\mathbf{S}\in\mcS_{k,m}^n}E\bigg[E\Big[\sum_{i=1}^{r+1}\sum_{j=\tau_{i-1}+k+1}^{\tau_i-k}\Lambda(X_j,s_{k,j}(Z_{j-k}^{j+k}))\Big|A^n\Big]\bigg]\label{th5 eq10}\\
&=&\frac{1}{n}E\bigg[\sum_{i=1}^{r+1}\sum_{j=\tau_{i-1}+k+1}^{\tau_i-k}\min_{s_k\in\mcS_k}E\Big[\Lambda(X_j,s_{k}(Z_{j-k}^{j+k}))\Big|A^n\Big]\bigg]\label{th5 eq5}\\
&=&\frac{1}{n}E\bigg[\sum_{i=1}^{r+1}\sum_{j=\tau_{i-1}+k+1}^{\tau_i-k}\sum_{z_{-k}^k\in\mcZ^{2k+1}}P(Z_{j-k}^{j+k}=z_{-k}^{k}|A^n)\min_{\hat{x}\in\hat{\mathcal{X}}}E\Big[\Lambda(X_j,\hat{x})|Z_{j-k}^{j+k}=z_{-k}^{k},A^n\Big]\bigg]\label{th5 eq6}\\
&=&\frac{1}{n}E\bigg[\sum_{i=1}^{r+1}\sum_{j=\tau_{i-1}+k+1}^{\tau_i-k}\sum_{z_{-k}^k\in\mcZ^{2k+1}}P(Z_{j-k}^{j+k}=z_{-k}^{k}|A^n)U_{\Lambda}(\mathbf{P}_{X_j|Z_{j-k}^{j+k}=z_{-k}^k,A^n})\bigg]\label{th5 eq7}\\
&=&\frac{1}{n}E\bigg[\sum_{i=1}^{r+1}\sum_{j=\tau_{i-1}+k+1}^{\tau_i-k}E\big[U_{\Lambda}(\mathbf{P}_{X_j|Z_{j-k}^{j+k},A^n})\big|A^n]\bigg]\nonumber\\
&=&\frac{1}{n}E\bigg[\sum_{i=1}^{r+1}\sum_{j=\tau_{i-1}+k+1}^{\tau_i-k}E\big[U_{\Lambda}(\mathbf{P}^{(A_{\tau_i})}_{X_0|Z_{-k}^{k}})|A^n\big]\bigg]\label{th5 eq9}\\
&\leq&\frac{1}{n}E\bigg[\sum_{i=1}^{r+1}\sum_{j=\tau_{i-1}+1}^{\tau_i}E\big[U_{\Lambda}(\mathbf{P}^{(A_{\tau_i})}_{X_0|Z_{-k}^{k}})|A^n\big]\bigg],\label{th5
eq11}
\end{eqnarray}
where \eq{th5 eq5} follows from the stationarity of the distribution
in each block as well as the fact that the combination of the best
$k$-th order sliding window denoiser for each block is in
$\mcS_{k,m}^n$ and achieves the minimum in \eq{th5 eq10}; \eq{th5
eq6} follows from conditioning; \eq{th5 eq7} follows from the
definition \eq{U def}; \eq{th5 eq9} follows from the stationarity of
the distribution in each $i$-th block; and \eq{th5 eq11} follows
from adding more nonnegative terms.

For the second term in (ii), we first define
$$
n_i(A^n)\triangleq\tau_i(A^n)-\tau_{i-1}(A^n)
$$
as the length of the $i$-th block, for $1\leq i\leq r(A^n)+1$.
Obviously, $n_i(A^n)$ also depends on $A^n$, and, thus, is a random
variable, but we again suppress $A^n$ for brevity and denote it as
$n_i$. Then, similar to the first term above, we obtain
\begin{eqnarray}
\mathbb{D}(P_{X^n},\mathbf{\Pi})&=
&\min_{\hat{\mathbf{X}}^n\in\mathcal{D}_n}EL_{\hat{\mathbf{X}}^n}(X^n,Z^n)\nonumber\\
&=&\frac{1}{n}\min_{\hat{\mathbf{X}}^n\in\mathcal{D}_n}E\bigg[E\Big[\sum_{i=1}^{r+1}\sum_{j=\tau_{i-1}+1}^{\tau_i}\Lambda(X_j,\hat{X}_j(Z^n))\Big|A^n\Big]\bigg]\nonumber\\
&=&\frac{1}{n}E\bigg[\sum_{i=1}^{r+1}\sum_{j=\tau_{i-1}+1}^{\tau_i}\min_{\hat{X}:\mcZ^n\rightarrow\mcXhat}E\Big[\Lambda(X_j,\hat{X}(Z^n))\Big|A^n\Big]\bigg]\nonumber\\
&=&\frac{1}{n}E\bigg[\sum_{i=1}^{r+1}\sum_{j=\tau_{i-1}+1}^{\tau_i}\min_{\hat{X}:\mcZ^{n_i}\rightarrow\mcXhat}E\Big[\Lambda(X_j,\hat{X}(Z_{\tau_{i-1}+1}^{\tau_i}))\Big|A^n\Big]\bigg]\label{th5 eq12}\\
&=&\frac{1}{n}E\bigg[\sum_{i=1}^{r+1}\sum_{j=\tau_{i-1}+1}^{\tau_i}E\big[U_{\Lambda}(\mathbf{P}_{X_j|Z_{\tau_{i-1}+1}^{\tau_i},A^n})\big|A^n\big]\bigg]\nonumber\\
&=&\frac{1}{n}E\bigg[\sum_{i=1}^{r+1}\sum_{j=\tau_{i-1}+1}^{\tau_i}E\big[U_{\Lambda}(\mathbf{P}^{(A_{\tau_i})}_{X_0|Z_{1-j}^{n_i-j}})\big|A^n\big]\bigg]\label{th5 eq13}\\
&\geq&\frac{1}{n}E\bigg[\sum_{i=1}^{r+1}\sum_{j=\tau_{i-1}+1}^{\tau_i}E\big[U_{\Lambda}(\mathbf{P}^{(A_{\tau_i})}_{X_0|Z_{-\infty}^{\infty}})\big|A^n\big]\bigg],\label{th5
eq14}
\end{eqnarray}
where \eq{th5 eq12} follows from the conditional independence
between different blocks, given $A^n$; \eq{th5 eq13} follows from
the stationarity of the distribution in each block, and \eq{th5
eq14} follows from \cite[Lemma 4(1)]{Dude}. Therefore, from
\eq{mineq1},\eq{th5 eq11}, and \eq{th5 eq14}, we obtain
\begin{eqnarray}
(b)&=&\min_{\mathbf{S}\in\mcS_{k,m}^n}EL_{\hat{\mathbf{X}}^{n,\mathbf{S}}}(X^n,Z^n)-\mathbb{D}(P_{X^n},\mathbf{\Pi})\nonumber\\
   &\leq&\frac{2k(m+1)\Lambda_{\max}}{n}+\frac{1}{n}E\bigg[\sum_{i=1}^{r+1}\sum_{j=\tau_{i-1}+1}^{\tau_i}E\big[U_{\Lambda}(\mathbf{P}^{(A_{\tau_i})}_{X_0|Z_{-k}^{k}})|A^n\big]-E\big[U_{\Lambda}(\mathbf{P}^{(A_{\tau_i})}_{X_0|Z_{-\infty}^{\infty}})\big|A^n\big]\bigg]\nonumber\\
   &=&\frac{2k(m+1)\Lambda_{\max}}{n}+E\bigg[\sum_{i=1}^{r+1}\frac{n_i}{n}\cdot\Big\{E\big[U_{\Lambda}(\mathbf{P}^{(A_{\tau_i})}_{X_0|Z_{-k}^{k}})|A^n\big]-E\big[U_{\Lambda}(\mathbf{P}^{(A_{\tau_i})}_{X_0|Z_{-\infty}^{\infty}})\big|A^n\big]\Big\}\bigg].\label{th5
   eq15}
\end{eqnarray}
Now, observe that, regardless of $A^n$, the sequence of numbers
$\{\frac{n_i}{n}\}_{i=1}^{r+1}$ form a probability distribution,
since $\sum_{i=1}^{r+1}\frac{n_i}{n}=1$ and $\frac{n_i}{n}\geq 0$
for all $i$, with probability 1. Then, based on the fact that the
average is less than the maximum,  we obtain the further upper bound
\begin{eqnarray}
\eq{th5
eq15}&\leq&\frac{2k(m+1)\Lambda_{\max}}{n}+E\bigg[\max_{i\in\{1,\cdots,M\}}\Big\{E\big[U_{\Lambda}(\mathbf{P}^{(i)}_{X_0|Z_{-k}^{k}})\big]-E\big[U_{\Lambda}(\mathbf{P}^{(i)}_{X_0|Z_{-\infty}^{\infty}})\big]\Big\}\bigg].\label{final
ub}
\end{eqnarray}

The remaining argument to prove the theorem is to show that the
upper bounds (\ref{th5 eq3}) and (\ref{final ub}) converge to 0 as
$n$ tends to infinity. First, from the given condition on $k=k_n$
and $m=m_n$, the maximal allowable growth rate for $k$ is $k=c_1\log
n$, which leads to
$\lim_{n\rightarrow\infty}\frac{2k\Lambda_{\max}}{n}=0$. In
addition, the condition requires $m=o(n)$, and $k$ to be
sufficiently slow, such that
$k|\mcZ|^{2k}=o\big((\frac{n}{m})^{1-\delta}\big)$, which implies
$k=o(\frac{n}{m})$. Therefore,
$\lim_{n\rightarrow\infty}\frac{2k(m+1)\Lambda_{\max}}{n}=0$.
Furthermore, from conditioning on $X^n$, bounded convergence
theorem, and part b) of Theorem \ref{semi stochastic theorem}, we
obtain
$\lim_{n\rightarrow\infty}E[L_{\hat{\mathbf{X}}_{\mathrm{univ}}^{n,k,m}}(X^{n-k}_{k+1},Z^n)-D_{k,m}(X^n,Z^n)]=0$.
Thus, we have
\begin{eqnarray}
&
&\limsup_{n\rightarrow\infty}\Big[EL_{\hat{\mathbf{X}}_{\mathrm{univ}}^{n,k,m}}(X^n,Z^n)-\mathbb{D}(P_{X^n},\mathbf{\Pi})\Big]\nonumber\\
&\leq&\limsup_{n\rightarrow\infty}E\bigg[\max_{i\in\{1,\cdots,M\}}\Big\{E\big[U_{\Lambda}(\mathbf{P}^{(i)}_{X_0|Z_{-k}^{k}})\big]-E\big[U_{\Lambda}(\mathbf{P}^{(i)}_{X_0|Z_{-\infty}^{\infty}})\big]\Big\}\bigg]\nonumber\\
&\leq&E\bigg[\limsup_{n\rightarrow\infty}\max_{i\in\{1,\cdots,M\}}\Big\{E\big[U_{\Lambda}(\mathbf{P}^{(i)}_{X_0|Z_{-k}^{k}})\big]-E\big[U_{\Lambda}(\mathbf{P}^{(i)}_{X_0|Z_{-\infty}^{\infty}})\big]\Big\}\bigg]\label{th5 eq16}\\
&=&0\label{th5 eq17},
\end{eqnarray}
where \eq{th5 eq16} follows from the reverse Fatou's lemma, and
\eq{th5 eq17} follows from \cite[Lemma 4(2)]{Dude} and $M$ being
finite. Since it is clear that $
\liminf_{n\rightarrow\infty}[EL_{\hat{\mathbf{X}}_{\mathrm{univ}}^{n,k,m}}(X^n,Z^n)-\mathbb{D}(P_{X^n},\mathbf{\Pi})]\geq
0 $ by definition of $\mathbb{D}(P_{X^n},\mathbf{\Pi})$, the theorem
is proved. $\quad\blacksquare$\\

\noindent \emph{Remark:} As in \cite[Theorem 3]{Dude}, the
convergence rate in \eq{eq: s-dude universality} may depend on
$P_\mathbf{X}$, and there is no vanishing upper bound on this rate
that holds for all $P_{\mathbf{X}}\in\mathcal{P}\{m_n\}$. However,
we can glean some insight into the convergence rate from (i) and
(ii): whereas the term (i) is uniformly upper bounded for all
$P_{\mathbf{X}}\in\mathcal{P}\{m_n\}$,\footnote{Recall part b) of
Theorem \ref{semi stochastic theorem}, where a uniform bound
(uniform in the underlying individual sequence) on $
E\big[L_{\hat{\Xb}^{n,k,m}_{\mathrm{univ}}}(x^n,Z^n)-D_{k,m}(x^n,Z^n)\big]$
was provided in the semi-stochastic setting. Clearly, in the
stochastic setting the same bound holds  on $
E\big[L_{\hat{\Xb}^{n,k,m}_{\mathrm{univ}}}(X^n,Z^n)-D_{k,m}(X^n,Z^n)\big]$,
regardless of the distribution of $X^n$.} the rate at which term
(ii) vanishes
 depends on $P_{\mathbf{X}}$. In general, we observe that the slower the rate of increase
 of $k=k_n$, the faster  the convergence in (i), but
the convergence in (ii) is slower. With respect to the rate of
increase of $m_n$, the slower it is, the faster the convergence in
(i), but whether or not the convergence in (ii) is accelerated by a
slower rate of increase of $m_n$ may depend on the underlying
process distribution $P_{\mathbf{X}}$.

\end{document}